\documentclass[
    prx, aps, superscriptaddress, twocolumn, longbibliography, nofootinbib
]{revtex4-2}

\usepackage[english]{babel}
\usepackage[T1]{fontenc}
\usepackage{microtype}
\usepackage{
    array, multirow, enumerate, enumitem,
    amsmath, amsfonts, amsthm, amssymb, mathtools, mathbbol, mathdots, thmtools, centernot, bm, physics,
    hyperref,
    graphicx, adjustbox, tikz
}
\usepackage{xurl}
\usepackage[capitalise]{cleveref}
\usepackage{ragged2e}
\usepackage[ruled, vlined]{algorithm2e}
\SetAlgoCaptionLayout{justify}
\usepackage{hhline}
\crefname{section}{Sec.}{Secs.}
\Crefname{section}{Sec.}{Secs.}
\crefname{subsection}{Sec.}{Secs.}
\Crefname{subsection}{Sec.}{Secs.}
\crefname{appendix}{App.}{Apps.}
\Crefname{appendix}{App.}{Apps.}
\crefname{subappendix}{App.}{Apps.}
\Crefname{subappendix}{App.}{Apps.}
\usetikzlibrary{quantikz2}
\hypersetup{
    colorlinks=true,
    linkcolor=blue,
    citecolor=blue,
    urlcolor=blue
}

\theoremstyle{plain}
\newtheorem{thm}{Theorem}

\theoremstyle{definition}

\newtheorem{prob}{Problem}

\newcommand{\lap}{\upsilon} 

\begin{document}

\title{Calculating response functions of coupled oscillators using quantum phase estimation}

\author{Sven Danz}
\email{s.danz@fz-juelich.de}
\affiliation{Institute for Quantum Computing Analytics (PGI-12), Forschungszentrum J\"ulich, 52425 J\"ulich, Germany}
\affiliation{Theoretical Physics, Saarland University, 66123 Saarbrücken, Germany}
\author{Mario Berta}
\affiliation{Institute for Quantum Information, RWTH Aachen University, 52074 Aachen, Germany}
\affiliation{Department of Computing, Imperial College London, London SW7 2RH, UK}
\author{Stefan Schröder}
\affiliation{Fraunhofer Institute for Production Technology IPT,  52074 Aachen, Germany}
\author{Pascal Kienast}
\affiliation{Fraunhofer Institute for Production Technology IPT,  52074 Aachen, Germany}
\author{Frank K. Wilhelm}
\affiliation{Institute for Quantum Computing Analytics (PGI-12), Forschungszentrum J\"ulich, 52425 J\"ulich, Germany}
\affiliation{Theoretical Physics, Saarland University, 66123 Saarbrücken, Germany}
\author{Alessandro Ciani}
\affiliation{Institute for Quantum Computing Analytics (PGI-12), Forschungszentrum J\"ulich, 52425 J\"ulich, Germany}

\begin{abstract}
We study the problem of estimating frequency response functions of systems of coupled, classical harmonic oscillators using a quantum computer. 
The functional form of these response functions can be mapped to a corresponding eigenproblem of a Hermitian matrix $H$, thus suggesting the use of quantum phase estimation.
Our proposed quantum algorithm operates in the standard $s$-sparse, oracle-based query access model.
For a network of $N$ oscillators with maximum norm $\lVert H \rVert_{\mathrm{max}}$, and when the eigenvalue tolerance $\varepsilon$ is much smaller than the minimum eigenvalue gap, we use $\mathcal{O}(\log(N s \lVert H \rVert_{\mathrm{max}}/\varepsilon)$ algorithmic qubits and obtain a rigorous worst-case query complexity upper bound $\mathcal{O}(s \lVert H \rVert_{\mathrm{max}}/(\delta^2 \varepsilon) )$ up to logarithmic factors, where $\delta$ denotes the desired precision on the coefficients appearing in the response functions. 
Crucially, our proposal does not suffer from the infamous state preparation bottleneck and can as such potentially achieve large quantum speedups compared to relevant classical methods.
As a proof-of-principle of exponential quantum speedup, we show that a simple adaptation of our algorithm solves the random glued-trees problem in polynomial time.
We discuss practical limitations as well as potential improvements for quantifying finite size, end-to-end complexities for application to relevant instances.
\end{abstract}

\maketitle


\section{Introduction}
\label{sec:intro}
It is part of the experience of scientists and engineers that many relevant problems appearing in nature can be formulated as the problem of finding the eigenvalues and eigenvectors of a matrix. Consequently, a natural question to ask is whether quantum computers can provide some advantages in solving eigenproblems, compared to standard classical algorithms. The answer to this question is the subject of active research, but it is highly nontrivial and depends on the fine details of the problem itself. 

In this paper, we focus on a particular category of eigenproblems, namely those associated with a system of $N$ coupled harmonic oscillators. 
Needless to say, systems of coupled harmonic oscillators are ubiquitous in nature, as they can be used to effectively model the low energy dynamics of apparently very different physical systems, from electrical circuits to the vibration of molecules. 
In essence, our goal is to perform a modal analysis of the system, focusing on the calculation of response functions in frequency (or Laplace) domain. 
Response functions describe how much an external perturbation applied to a certain oscillator, influences the dynamics of another oscillator in the system. 
These functions have an analytical functional form that is related to an eigenproblem defined on the system of coupled oscillators. We thus focus on the question of how these response functions can be extracted using a quantum algorithm and in which cases we might expect significant quantum speedups.


\subsection{Related work}
\label{subsec:prevwork} 
The idea of using a quantum computer for eigenproblems dates back to the early days of quantum algorithm development \cite{abrams1999}. 
In the quantum chemistry setting, where one is typically interested in characterizing the low-lying eigenvalues and eigenstates of a molecular Hamiltonian, it is in fact considered one of the most promising applications of quantum computers \cite{Bauer2020, mcardle2020, dalzell2023}. 

Broadly speaking, one can distinguish between two categories of quantum algorithms for eigenproblems, namely algorithms based on the variational quantum eigensolver \cite{Peruzzo2014, TILLY20221}, which lack rigorous scaling guarantees, and those based on the quantum phase estimation (QPE) subroutine \cite{NielsenChuang, kitaevBook, cleve1998}. 
In this work, we focus on the latter, and in particular on a formulation of standard QPE for eigenvalue estimation that relies on the block-encoding of the matrix of interest in a (larger) unitary \cite{Low2019hamiltonian, low2017, gilyen2019, martyn2021, tang2023, dalzell2023, poulin2018, berry2018}. 
A key feature of the problem that we exploit is that to estimate the response functions, we do not need to invoke a state preparation subroutine that prepares a desired eigenvector as a quantum state, for instance via controlled rotations~\cite{grover2002}, reject sampling method~\cite{ozols2012}, eigenvalue transform~\cite{macardle2022} or adiabatic state preparation (see discussion in \cite{dalzell2023}). These kinds of state preparation subroutines hide additional complexity and can ruin potential quantum speedups when properly taken into account. Instead, as we will show, in our case the omnipresent state preparation problem in QPE reduces to the preparation of a product state, from which we need to determine the relevant eigenvalues and associated weights via importance sampling. The simultaneous estimation of multiple eigenvalues using either QPE or other quantum subroutines has been studied in the literature using different approaches \cite{obrien_2019, dutkiewicz2022, ding2023, Somma_2019}.  

The works that are most related to ours on resolving coupled harmonic oscillators on a quantum computers are Refs.~\cite{babbush2023,lee2023}. Ref.~\cite{babbush2023} analyzes the problem of obtaining the classical time-evolution of a system of coupled harmonic oscillators. To do this, the authors employ a particular encoding of the positions and momenta of the oscillators in the amplitudes of an $n$-qubit quantum system. On this system, a Hamiltonian can be defined that mimics the time-evolution of the system of coupled oscillators at any time. With this approach, while one cannot efficiently access the full solution, it is possible to estimate efficiently relevant quantities such as the kinetic energy of a subset of oscillators. Moreover, Ref.~\cite{babbush2023} also shows that the random glued-trees problem originally studied in Ref.~\cite{childs2003} can be mapped to a problem on coupled oscillators and solved efficiently using the quantum approach. We complement this result by showing that the glued-trees problem can also be solved efficiently with our QPE approach by a simple adaptation of the algorithm for the estimation of response functions. An interesting open question is whether response functions can also be estimated using the time-evolution approach of Ref.~\cite{babbush2023}. 

Ref.~\cite{lee2023} instead focuses on a normal mode analysis of systems of coupled oscillators, in the same spirit as our work, and essentially using the same kind of embedding of the problem in quantum systems. However, Ref.~\cite{lee2023} does not discuss the application to the estimation of response functions and assumes that a good approximation of the eigenvectors can be prepared in a quantum system. While this would also allow to estimate response functions in principle, we show that it is not necessary for this task. 

Finally, our work is also connected to quantum approaches to solve the wave equation \cite{costa2019} or more generally for finite element simulations of solid structures \cite{clader2013, montanaro2016}. In fact, as we detail in the next section, once discretized, these problems can be mapped to a system of coupled harmonic oscillators. 

    \subsection{Background on applications}
    \label{subsec:motivation}

The solution of ordinary and partial differential or integral equations describing technical and physical processes, such as the deformation of components under given loads or heat conduction in solids, is rarely successful analytically and must be solved numerically. 
The necessary transfer to a finite-dimensional equivalent problem requires a discretization process using methods such as the finite element or finite volume method~\cite{jung13}. 
In our work, we are particularly interested in problems for which this discretization step leads to an  effective model of coupled harmonic oscillators. 
A prominent example of these problems are those that emerge in the manufacturing industry as we detail below. 

\begin{figure}
    \centering
    \includegraphics[width=0.9\linewidth]{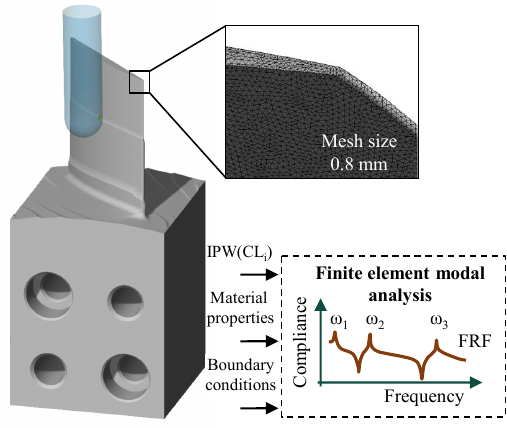}
    \caption{
        In-process-workpiece (IPW) of a single blade demonstrator (left) and the detailed representation of the finite element mesh. The cutting tool is shown in blue. With the cutter location-dependent IPW, material properties (density, Young's modulus, Poisson's ratio), and boundary conditions as input, the finite element modal analysis can be conducted and the frequency response function (FRF) calculated by solving the corresponding eigenproblem. Typical solvers used for this task are based on the QR algorithm and have complexity $\mathcal{O}(N^3)$ \cite{trefethen97, press1992, arbenzNotes}.  
    }
    \label{fig:femmesh}
\end{figure}

Due to their time efficiency, finite element models are often used in practice to simulate machining processes, such as milling. Milling is a metal cutting manufacturing process that uses the circular cutting motion of a tool, usually with multiple cutting edges, to produce a vast variety of surfaces on a workpiece (see \cref{fig:femmesh}).
In all milling processes, in contrast to other processes, such as turning and drilling for instance, the cutting edges are not constantly engaged. Still, at least one cutting interruption per cutting edge occurs with each revolution of the tool~\cite{klocke2018f}.   
Due to the constant cutting interruptions, depending on the milling cutter speed, a dynamic excitation of the workpiece and the tool can occur, which can have a negative effect on the surface quality in the form of vibration marks. 
Dynamic process stability simulations are carried out to analyze the vibrations and to improve the process design for the milling of thin-walled components~\cite{25_Maslo_S, munoa2016, quintana2011, budak2012, altintas2008}. The ultimate goal of these simulations is to obtain response functions over a certain frequency range, i.e., the typical frequency range of the cutting tool, that provide information about the effect that the cutting tool at a certain location has on the workpiece at other, possibly different, locations. For this particular case, the response function is a so-called compliance (with units $\mathrm{m}/\mathrm{N}$) as a function of the frequency of the cutting edge, as shown in \cref{fig:femmesh}, which effectively quantifies the vibration of the workpiece caused by the cutting tool. 

Ref.~\cite{schroeder24} analyzed a workflow for milling dynamics simulation, examining computational time and identifying numerical problems that can be suited for quantum algorithms. The milling dynamics workflow considers the dynamic behavior of the workpiece, which continuously varies due to changes in the stiffness and mass of the workpiece caused by the material removal and the position-dependent force excitation by the milling tool~\cite{rudel_2022}.
This analysis highlighted the fact that finite element modal analysis is typically a computationally intensive simulation application in milling of thin-walled aerospace components, and as such could benefit from possible advantages that quantum algorithms could offer for this task.

    \subsection{Structure}
    \label{subsec:structure}

The rest of the paper is structured as follows.
In \cref{sec:discuss}, we discuss the connection between the computation of the response function of coupled oscillators and the solution of an eigenproblem. We further describe briefly our quantum algorithm and its scaling behavior, which is summarized in \cref{thm:mainthm}.
We finish the section with a critical assessment of the limitations and possible improvements of our quantum approach. 
In \cref{sec:algo}, we describe our proposed quantum algorithm in more detail, for which a  pseudocode is provided in Algorithm~\ref{alg:resp_func}. \cref{sec:glued_tree} discusses how the random-glued trees problem can be solved using the QPE subroutine. 
We then summarize our results and provide an outlook in \cref{sec:conclusion}.
The appendices contain a more in-depth derivations of the analytical form of the response function (\cref{app:responseform}), descriptions of the quantum subroutines  used in the algorithm (\cref{app:quantumsubroutines}), a detailed error analysis (\cref{app:detail_error}), a modified version of the algorithm for nonlocal response functions (\cref{app:nonlocal_resp}), and a scaling analysis of the algorithm applied to the random glued-trees problem (\cref{app:gluedtree}).

\section{Problem statement and main results}
\label{sec:discuss} 

	\subsection{Linear response functions as an eigenproblem}
	\label{subsec:problstate}

\begin{figure}
    \centering
    \includegraphics[width= 5cm]{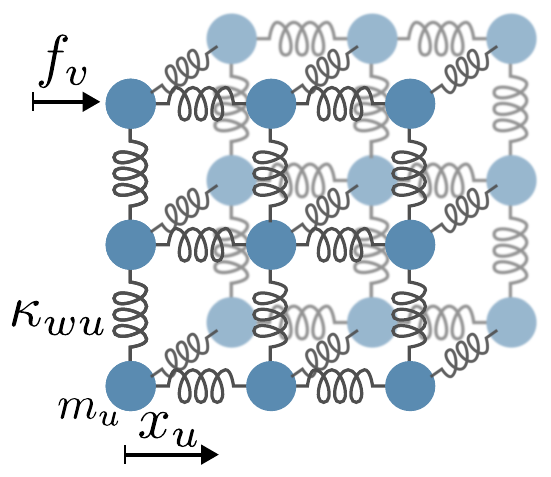}
    \caption{
        A representation of coupled oscillators. Point-like masses $m_u$ in blue are connected by springs with spring constant $\kappa_{wu}$. The force $f_v$ applied to mass~$v$ yields a displacement $x_u$ at mass~$u$.
    }
    \label{fig:coupledosc}
\end{figure}
Let us consider a system of masses coupled by spring constants as shown in Fig.~\ref{fig:coupledosc} and let $\mathcal{G} = (\mathcal{V}, \mathcal{E})$ be a graph with $\mathcal{V}$ the set of vertices and $\mathcal{E}$ the set of edges, that describes the network of oscillators. With each vertex $u \in \mathcal{V}$ we associate a harmonic oscillator with mass $m_u > 0$ and a coupling to a common wall with spring constant $\kappa_{u} >0$.
We denote by $N = \lvert \mathcal{V} \rvert$ the number of oscillators. With each edge $(u, v) \in \mathcal{E}$ in the graph, instead, we associate a coupling spring constant 
$\kappa_{u v}>0$. In what follows, we implicitly assume an ordering of the vertices, i.e., we associate with each $u \in \mathcal{V}$ a number between $1$ and $N$. 
For notational simplicity, we  also denote this number by $u$, since its meaning is clear from the context. Additionally,  we denote by $N_\mathcal{G}[u]$ the closed neighborhood of $u \in \mathcal{V}$, i.e., the set of vertices connected to $u$ by an edge plus $u$ itself. We call $s_u = |N_\mathcal{G}[u]|$ and define the sparsity $s$ as $s = \mathrm{max}_{u \in \mathcal{V}} s_u$. For later convenience, we also define the set of vertices $\tilde{N}_\mathcal{G}[u]$ such that $N_\mathcal{G}[u] \subset \tilde{N}_\mathcal{G}[u]$ and $|\tilde{N}_\mathcal{G}[u]| = s$. Basically, $\tilde{N}_\mathcal{G}[u]$ is the neighborhood of $u$ enlarged with dummy vertices to ensure that the total number of elements in the set is $s$. 

\subsubsection{Unperturbed dynamics}
We can compactly write the dynamical equations that govern the motion of the coupled oscillator systems, by defining two $N \times N$ matrices, namely a diagonal mass matrix $\bm{M}$ with matrix elements
\begin{equation}
M_{u v} = m_u \delta_{uv},
\end{equation}
and a stiffness matrix $\bm{K}$ with matrix elements
\begin{equation}
K_{u v} = \begin{cases} 
\kappa_u + \sum_{w \in N_\mathcal{G}[u] \setminus \{u\}} \kappa_{w u} \quad & \mathrm{if} \quad u=v, \\
-\kappa_{uv} \quad & \mathrm{if} \quad (u, v) \in \mathcal{E}, \\
0 \quad & \mathrm{otherwise}.
\end{cases}
\end{equation}
Additionally, we denote by $\bm{x}$ the $N$-dimensional column vector whose elements are the oscillator positions $x_u$ with respect to the equilibrium state.

The system evolves according to the dynamical equation
\begin{equation}
\label{eq:dyneq1}
    \bm{M} \frac{d^2 \bm{x}}{dt^2} = - \bm{K} \bm{x}. 
\end{equation}
To transform this problem into an eigenproblem, let us first define the matrix
\begin{equation}
\label{eq:hamiltonian}
    \bm{H} = \bm{M}^{-1/2} \bm{K} \bm{M}^{-1/2},
\end{equation}
which, as we will see, will play the role of a Hamiltonian in our quantum algorithm. 
Note that since $\bm{M}$ is positive definite, also $\sqrt{\bm{M}}$ is positive definite and thus invertible.
The matrix $\bm{H}$ is Hermitian and positive-semidefinite and we denote its eigenvalues by $\lambda_j \ge 0$ for $j\in\{1, \dots, N\}$.
We now show that solving the dynamical equation \cref{eq:dyneq1} is equivalent to diagonalizing the matrix $\bm{H}$.  
We denote by $\bm{W}$ an orthogonal matrix that diagonalizes $\bm{H}$,
\begin{equation}
\bm{\Lambda} = \bm{W}^T \bm{H} \bm{W},
\end{equation}
where $\bm{\Lambda}$ is a diagonal matrix with the eigenvalues of $\bm{H}$ on the diagonal, i.e., $\Lambda_{jj} = \lambda_j$. 
Defining the vector of normal mode variables $\bm{y}$ as
\begin{equation}
\label{eq:yvar}
\bm{y} = \bm{W}^T \bm{M}^{1/2} \bm{x},
\end{equation} 
we get an uncoupled system of harmonic oscillators satisfying
\begin{equation}
\frac{d^2 \bm{y}}{dt^2} = -\bm{\Lambda} \bm{y} \implies  \frac{d^2 y_j}{dt^2} = - \lambda_j y_j.
\end{equation}
Since $\lambda_j\ge 0$ they are usually denoted as $\lambda_j = \omega_j^2$, with $\omega_j$ the resonance frequency of the normal mode $j$. Note that \cref{eq:yvar} implies that 
\begin{equation}
\label{eq:xwujy}
    x_u = \frac{1}{\sqrt{m_u}}\sum_{j =1}^N W_{u j} y_j.
\end{equation}

\subsubsection{Adding an external force}
Let us now add an external force to Eq.~\eqref{eq:dyneq1}.  In particular, we assume that a time-dependent force $f_v(t)$ is applied to the oscillator $v \in \mathcal{V}$ and we gather all the forces in an $N$-dimensional column vector $\bm{f}(t)$. Including the forcing term, the dynamical equation reads
\begin{equation}
\label{eq:dyneqf}
    \bm{M} \frac{d^2 \bm{x}}{dt^2} = - \bm{K} \bm{x} + \bm{f}. 
\end{equation}
Eq.~\eqref{eq:dyneqf} gives rise to a convolution relation between $\bm{f}(t)$ and $\bm{x}(t)$ 
\begin{equation}
\label{eq:lti}
    \bm{x}(t) = \int_{- \infty}^{+\infty} d \tau \bm{g}(t - \tau) \bm{f}(\tau),
\end{equation}
where $\bm{g}(t)$ is an $N \times N$ matrix that we call the matrix response function. 
The matrix elements $g_{u v}(t)$ can simply be interpreted as the response of the oscillator $u$ when a forcing term on the oscillator $v$, $f_v(t)$, is a Dirac delta function, i.e., $f_v(t) = \delta(t)$, while no force is applied to the other oscillators.\footnote{The response function is also called the Green's function for the system of coupled oscillators.}
This interpretation also suggests that causality implies $\bm{g}(t) = 0$ if $t <0$ (i.e we consider a retarded response function). 
Thus, we can simply take the upper limit of the integral in \cref{eq:lti} to be $t$ rather than $+ \infty$. 
\cref{eq:lti} is the defining equation of a linear time-invariant system~\cite{triverio2007}. 

Now, let us introduce the bilateral Laplace transform of a generic function $\bm{z}(t)$ as
\begin{equation}
\label{eq:lapl}
\bm{Z}(\lap) = \int_{-\infty}^{+\infty} dt e^{- \lap t} \bm{z}(t), \quad \lap = \sigma + i \omega \in \mathbb{C}. 
\end{equation}
The Laplace transform is defined only in the region of convergence, that is the complex domain where the integral in Eq.~\eqref{eq:lapl} converges. 
From the convolution theorem for the Laplace transform Eq.~\eqref{eq:lti} in Laplace domain becomes
\begin{equation}
\bm{X}(\lap) = \bm{G} (\lap) \bm{F}(\lap),
\label{eq:response}
\end{equation} 
with $\bm{G}(\lap)$ the Laplace transform of the response function in time domain $\bm{g}(t)$, while $\bm{X}(\lap)$ and $\bm{F}(\lap)$ the Laplace transform of the position vector $\bm{x}(t)$ and the vector of forces $\bm{f}(t)$, respectively.

In the main text, we focus on a quantum algorithm to determine the diagonal elements of $\bm{G}(\lap)$ that take the form
\begin{equation}
\label{eq:diagresponse}
    G_{uu}(\lap) = \frac{1}{m_u}\sum_{j=1}^N \frac{W_{u j}^2}{\lambda_j + \lap^2},
\end{equation}
which we call the local response functions. However, it is also possible to obtain the off-diagonal elements of $\bm{G}(\lap)$ or nonlocal response function at the price of adding a Hadamard test, as we show in Appendix~\ref{app:nonlocal_resp}. 
We provide a derivation of \cref{eq:diagresponse}, as well as the more general definition for local and nonlocal response, in \cref{app:responseform}. 

\cref{eq:diagresponse} shows that the response function $G_{uu}(\lap)$ is completely determined once we solve the eigenproblem associated with $\bm{H}$. 
In fact, all we need to compute the response function is the eigenvalues $\lambda_j$ and the coefficients $W_{uj}^2$, with $W_{uj}$ the matrix elements of the orthogonal matrix $\bm{W}$ that diagonalizes $\bm{H}$. 
As discussed in \cref{subsec:motivation}, the response function is the relevant quantity to study in practical applications. 
State-of-the-art classical algorithms that are used in practice for the eigenproblem at hand, such as the QR algorithm, need at least $\mathcal{O}(N^3)$ arithmetic operations. 
In fact, most of the fastest classical algorithms for eigenvalue computation rely on the Householder transformation which transforms an $N\times N$ matrix into the Hessenberg form (almost triangular). 
This transformation needs $4N^3/3$ operations (see Chap. 11 p.~474 in~\cite{press1992})
and dominates the scaling of the QR algorithm.
The sparsity of $\bm{H}$ reduces the runtime to $\mathcal{O}(N^2s)$.

The quantum algorithm proposed in this manuscript is designed to compute all eigenpairs of a sparse matrix.
However, we will give a bief outlook on the scaling of advanced classical eigenpair solvers that compute only a small subset of all eigenpairs.
Both the power and the more advanced Lanczos method~\cite{Lanczos1950, ojalvo1970} allows us to approximate the $k$ largest eigenvalues of $\bm{H}$ in $\mathcal{O}(Nk(s+k))$ arithmetic operations if $\bm{H}$ has well-separated eigenpairs \cite{Paige1971,demmel1997}.
Combined with the inverse power method and a shifted matrix, this yields an approximation to an eigenvalue close to any chosen point~\cite{ericsson1980}.
However, the response function computed with those methods are only accurate at extremes of the full spectrum.
Its runtime scales worse for matrices with small eigenvalue gaps and its memory requirements with $\mathcal{O}(N(s+k))$ storage registers is also not small \cite{Paige1971}.

In the following quantum treatment, we refer to the $\bm{H}$ matrix in \cref{eq:hamiltonian} as the Hamiltonian. From now on we avoid using the bold symbols, simply denoting the Hamiltonian as $H$, and do the same for the other matrices.
	
	\subsection{Main idea}
	\label{subsec:mainres}
In this work, we describe a quantum algorithm that computes the local and nonlocal response functions of a system of coupled oscillators. 
We start by providing the basic intuition behind the algorithm for the case of the local response function in \cref{eq:diagresponse}, while we refer to \cref{sec:algo} and the appendices for more technical details. 
To compute $G_{uu}(\lap)$, we need to determine the eigenvalues $\lambda_j$ and the coefficients $W_{u j}^2$. 
In our approach, each vertex $u \in \mathcal{V}$ is associated with a bitstring $\ket{u}$ of length $n$ in the computational basis, where the number of qubits is $n = \lceil \log_2 N \rceil$. For simplicity, from now on we assume that $N$ is a power of two, i.e., $N=2^n$.
Since, $W$ is the orthogonal transformation that diagonalizes $H$, we can express the bitstring $\ket{u}$ as a linear combination of the eigenstates $\ket{\lambda_j}$ of $H$:
\begin{equation}
    \ket{u} 
    = 
    \sum_{j=1}^N 
        W_{u j} 
        \ket{\lambda_j}.
    \label{eq:vtolambda}
\end{equation}
Thus, we can interpret $W_{uj}$ as weights in the previous decomposition. 
Eqs.~\eqref{eq:diagresponse} and ~\eqref{eq:vtolambda} convey the core idea behind the algorithm. 
If we prepare the bitstring $\ket{u}$ and perform measurement in the eigenbasis $\ket{\lambda_j}$ of $H$ we will get the eigenvalues $\lambda_j $ with probability $W_{u j}^2$. 
By repeating this procedure many times we can estimate the coefficients $W_{u j}^2$ via statistical averages and obtain an estimate for the response function $G_{uu}(\lap)$. 
Note that there is an implicit importance sampling in this procedure: the eigenstates that contribute the most to $G_{uu}(\lap)$ are also those whose eigenvalue will be measured more often. 
To implement the measurement in the eigenbasis we use QPE.
The full algorithm is described in detail in Sec.~\ref{sec:algo}.

For the implementation of the quantum algorithm, we assume that we have oracle access to the matrix elements of $H$ and that $H$ is $s$-sparse, i.e., that each row has at most $s$ non-zero elements. In particular, we assume that we have access to two matrix oracles. 
The first one is the position oracle $O_P$, that acts on the $2n$-qubit state $\ket{u} \ket{0}^{\otimes n}$ as
\begin{equation}
    O_P\ket{u}\ket{0}^{\otimes n}
    =
    \frac{1}{\sqrt{s}}\sum_{v \in \tilde{N}_\mathcal{G}[u]}\ket{u}\ket{v}.
    \label{eq:O_P}
\end{equation}
$O_P$ stores in the second register a linear superposition of all the column indices that correspond to non-zero matrix elements at row $u$ and the additional dummy indices in $\tilde{N}_\mathcal{G}[u]$ if necessary.   

The second oracle is the angle oracle $O_\vartheta$ which contains information about the matrix elements of $H$. 
It encodes the angle 
\begin{equation}
\label{eq:anglehuv}
\vartheta_{uv}=\arccos\sqrt{\frac{\abs{H_{uv}}}{\norm{H}_\text{max}}},
\end{equation}
and the sign of $H_{uv}$ in an $(r+1)$-qubit ancilla register
\begin{equation}
    O_\vartheta\ket{u,v}\ket{0}^{\otimes r+1}
    =
    \ket{u,v}\ket{\vartheta_{uv},\Theta(-H_{uv})},
    \label{eq:angle_oracle}
\end{equation}
with $\Theta(\cdot)$ the Heaviside step function and $r$ the number of bits used to represent $\vartheta_{uv}$.
We renormalize $H_{uv}$ in \cref{eq:anglehuv} with $\norm{H}_\mathrm{max} = \max_{u,v}\abs{H_{uv}}$.
Specifically, since $0 \le \vartheta_{uv} \le \pi/2$, we assume the following $r$-bit representation of $\vartheta_{uv}$ 
\begin{equation}
\label{eq:rbittheta}
    \vartheta_{uv} = \frac{\pi}{2} \sum_{l=1}^r \vartheta_{uv}^{(l)} 2^{-l},
\end{equation}
with $\vartheta_{uv}^{(l)}\in\{0,1\}$. Thus, when we write $\ket{\vartheta_{uv}}$ in an $r$-qubit register, this means
\begin{equation}
\label{eq:rbitthetaket}
    \ket{\vartheta_{uv}} = \ket{\vartheta_{uv}^{(1)}} \ket{\vartheta_{uv}^{(2)}} \dots \ket{\vartheta_{uv}^{(r)}}. 
\end{equation}

Finally, we assume that we have a quantum adder $\mathrm{ADD}$ available, which adds the information in one quantum register to another, defined as
\begin{equation}
    \text{ADD}\ket{a}\ket{b}
    =
    \ket{a}\ket{b+a\mod 2^n},
\end{equation}
where $n$ is the number of qubits in the second register.\footnote{The number of qubits in the first register does not need to be equal to the one in the second register.}
The inverse of this operation is a quantum subtractor $\text{ADD}^\dagger = \text{SUB}$. Circuit realizations of the quantum adder are well known in the literature~\cite{Ruiz-Perez2017, cuccaro2004, draper2000, vedral1996}.
Together with the matrix oracles $O_P$ and $O_{\vartheta}$, the quantum adder $\mathrm{ADD}$ is needed for the implementation of the block-encoding described in detail in \cref{app:quantumsubroutines} (see \cref{fig:state_trafo} in particular) and used in \cref{sec:algo}.

The computational cost of the quantum adder will be neglected in the following query complexity analysis.
This is because the implementation of the oracle $O_\vartheta$ usually requires multiple queries to the quantum adder.
Hence, the additional use of one ADD per oracle is negligible in terms of required qubit number and computation time.

\begin{table}[]
    \centering
    \caption{Parameters used in the complexity analysis.}
    \label{tab:parameters}
    \renewcommand{\arraystretch}{1.5}
    \begin{tabular}{l|l}
        \hline 
        Symbol & Meaning\\
        \hline
        $N$ & Number of rows or columns of $H$ \\
        $s$ & Sparsity of $H$ \\
        $\norm{H}_\mathrm{max}$ & Maximum absolute value of the entries of $H$\\
        $\Delta_\lambda^{(u)}$ & Minimum eigenvalue gap at oscillator $u$\\
        $\varepsilon$ & Additive eigenvalue tolerance\\
        $\delta$ & Additive weight tolerance\\
        $\zeta$ & Probability of failure\\
        $N_u$ & Number of nonzero $W_{uj}$ at oscillator $u$ \\
        \hline
    \end{tabular}
\end{table}

\subsection{Complexity analysis}
\label{subsec:complexity}
We now summarize the main results in terms of scaling with respect to relevant parameters. We start by defining the additional parameters that enter in the scaling analysis. First, we define the minimum, positive gap in the eigenvalue spectrum as
\begin{equation}
    \Delta_{\lambda} = \mathrm{min}\{|\lambda_i - \lambda_j| : i, j\in\{1, \dots, N\}, \,|\lambda_i - \lambda_j| >0 \}.
\end{equation}
Additionally, we define the minimum, positive eigenvalue gap at oscillator $u$ as
\begin{multline}
\label{eq:deltalambdau}
    \Delta_{\lambda}^{(u)} = \mathrm{min}\{|\lambda_i - \lambda_j| : i, j\in\{1, \dots, N\}, \\
    \,|\lambda_i - \lambda_j| >0, W_{ui}^2 >0,W_{u j}^2 >0 \},
\end{multline}
with $W_{ui}$ and $W_{uj}$ being the coefficients in \cref{eq:vtolambda}. 
This is the parameter that matters in our algorithm since, in the presence of degeneracies, weights associated with the same eigenvalue would just be gathered together in \cref{eq:diagresponse} and eigenvectors that have no support at oscillator $u$ give zero contribution. Obviously, $\Delta_{\lambda}^{(u)} \ge \Delta_{\lambda}$.
The scaling further depends on the required, additive tolerances $\varepsilon$ and $\delta$ for the eigenvalues $\lambda_j$ and weights $W_{u j}^2$, respectively, which we treat as independent parameters of our choice \footnote{The smaller the error on the eigenvalues, the smaller the error on the weights, and vice versa. However, for our analysis it is easier to think about them as independent tolerances.}. Moreover, we define the parameter $N_u$ to be the number of eigenvectors that have support at oscillator $u$, that is for which $W_{uj}^2 >0$. Note that in the worst case scenario $N_u = N$ from the true value. Finally, we denote by $\zeta$ the probability that the estimate of $W_{uj}^2$ deviates by more than $\delta$. All the parameters used in the complexity analysis are summarized in \cref{tab:parameters}.

Our main result is described by the following theorem. 
\begin{thm}
\label{thm:mainthm}
Let $H$ be a $N \times N$ Hermitian matrix that describes a system of coupled oscillators as in \cref{eq:hamiltonian} and let $\mathcal{G}=(\mathcal{V}, \mathcal{E})$ be the graph associated with the network of oscillators. Let $H$ be $s$-sparse with maximum norm $\norm{H}_{\mathrm{max}}$. Given an oscillator $u \in \mathcal{V}$, there exists a quantum algorithm that approximates the eigenvalues $\lambda_j$ of $H$ and the weights $W_{uj}^2$ in \cref{eq:diagresponse}, with additive error $\varepsilon$ and $\delta$, respectively, that requires
\begin{equation}
\label{eq:ntotscaling}
    n_{\mathrm{tot}} = \mathcal{O}\left(
        \log_2\left(
            Ns \norm{H}_\text{max} 
            \max\left(
                \frac{1}{\varepsilon},
                \frac{1}{\delta\Delta_\lambda^{(u)}}
            \right)
        \right)
    \right)
\end{equation}
qubits and 
\begin{equation}
\label{eq:nquerytotscaling}
    N_{\mathrm{queries}}^{(\mathrm{tot})} 
    = 
    \mathcal{O} \Biggl(
        \frac{s \norm{H}_{\mathrm{max}}}{\delta^2} 
        \ln \left(
            \frac{N_u}{\zeta}    
        \right) 
        \max\left(
            \frac{1}{\varepsilon},
            \frac{1}{\delta\Delta_\lambda^{(u)}}
        \right) 
    \Biggl)
\end{equation}
total queries of the matrix oracles $O_P$ and $O_{\vartheta}$ in \cref{eq:O_P} and \cref{eq:angle_oracle}, respectively.  
\end{thm}

A detailed error analysis of the QPE-based algorithm behind \cref{thm:mainthm} is presented in \cref{app:detail_error}.
In what follows, we provide further comments on the complexity of our result. 

First, we note that \cref{eq:ntotscaling} implies that $\varepsilon$ should be taken at least $\varepsilon < \Delta_{\lambda}^{(u)}$, since $\delta < 1$, to use the full potential of $n_\mathrm{tot}$ qubits. 
Moreover, in \cref{eq:ntotscaling}, we are not considering the ancilla qubits necessary to implement the oracles $O_P$, $O_\vartheta$, since these are problem-dependent. 
If we use $r$ bits to represent the matrix elements, the required number of bits in classical algorithms is usually $\mathcal{O}(rN^2)$, while for $s$-sparse matrices it can be $\mathcal{O}(rsN)$ in the best case.
The gate complexity of our quantum algorithm depends on the complexity of implementing the oracles $O_P$ and $O_\vartheta$, which are, as mentioned above, problem-dependent. 
Generally, as long as the matrix elements of a matrix can be accessed efficiently classically, this should also hold on a quantum computer, although at the price of having several ancillas to make the computation reversible. 
As an example, for the case of a linear chain of oscillators, we provide a full compilation of the oracles into elementary gates in the code available in the repository associated with this manuscript (see the ``Data Availability" section). 
As such, we choose to present the complexity of our algorithm in terms of the number of oracle queries $N_{\mathrm{queries}}$. 
The query complexity of a single run of QPE is (see \cref{subapp:oraclequeries})
\begin{equation}
\label{eq:queryscaling}
	N_{\mathrm{queries}} = \mathcal{O}\left(
        s \norm{H}_\text{max} \max\left(
            \frac{1}{\varepsilon},
            \frac{1}{\delta\Delta_\lambda^{(u)}}
        \right)
	\right).
\end{equation}
This is mostly due to the repetition of the controlled $V$ operator in the core of the phase estimation (cf. \cref{fig:full_phase_estimation} and Algorithm~\ref{alg:resp_func}).

QPE is a probabilistic algorithm from which we sample multiple times to estimate the eigenvalues and the associated probabilities. As we show in \cref{subapp:sample_size}, the number of samples $N_\mathrm{S}$ needed to estimate each weight $W_{uj}^2$ via empirical averages with error $\delta$ and failure probability $\zeta$ scales as
\begin{equation}
\label{eq:samplescaling}
    N_\mathrm{S} = \mathcal{O}\left(
        \frac{1}{\delta^2}
        \ln\left(
            \frac{N_u}{\zeta}    
        \right)
    \right).
\end{equation}
In the worst case scenario when $N_u = N =2^n$, the scaling would be logarithmic with the size of the matrix $N$, i.e., linear with the number of qubits $n$. 
We highlight the fact that this is not equivalent to requiring that the probability distribution $W_{uj}^2$ is estimated with error $\delta$ in total variation distance. 
Using the definition of total variation distance and Hoeffding's inequality similarly to \cref{subapp:sample_size}, one can show that in this case the scaling would be linear, and not logarithmic, with $N_u$. 
Moreover, we are not requiring to find all the $N_u$ eigenvalues that contribute in the response function at oscillator $u$. 
This would require a number of samples scaling at least as $N_u$, but what matters for us is the evaluation of the weights $W_{uj}^2$ with error $\delta$, since the weights appear at the numerator in \cref{eq:diagresponse}. 
This leads to the inverse quadratic scaling in $\delta$ in \cref{eq:samplescaling}.

Combining \cref{eq:queryscaling,eq:samplescaling}, we obtain the total number of queries as in \cref{eq:nquerytotscaling}.

	\subsection{Critical assessment}
	\label{subsec:algolimit}
\subsubsection{Limitations of the algorithm}
Neglecting $\delta$ for a moment, the computational cost (cf. \cref{eq:ntotscaling,eq:nquerytotscaling}) contains the inverse of
\begin{equation}
    \frac{\min\left(
        \varepsilon,
        \Delta_\lambda^{(u)}
    \right)}{
    s\norm{H}_\mathrm{max}
    }.
\end{equation}
Assuming the worst-case scenario in which, $\Delta_{\lambda}^{(u)} = \Delta_{\lambda}$, and we want to guaranty a separation of all eigenvalues, the  desired tolerance $\varepsilon$ is required to satisfy $\varepsilon < \Delta_\lambda$.
We further know that the $1$-norm $\norm{H}_1=\max_v\sum_{u}\abs{H_{uv}}$ is smaller than $s\norm{H}_\mathrm{max}$. Thus, we have
\begin{equation}
    \frac{\min\left(
        \varepsilon,
        \Delta_\lambda
    \right)
    }{
        s\norm{H}_\mathrm{max}
    } 
    \leq
    \frac{\Delta_\lambda}{s\norm{H}_\mathrm{max}}
    \leq
    \frac{\Delta_\lambda}{\norm{H}_\mathrm{1}}
    \leq
    \frac{\Delta_\lambda}{\lambda_\mathrm{max}},
\end{equation}
where we used the upper limit for the maximum eigenvalue $\lambda_\mathrm{max}\leq\norm{H}_1$.
Hence, the number of oracle queries scales with the ratio between the smallest eigenvalue gap and the maximum eigenvalue.
This, in the general case, increases polynomially with the number of eigenvalues that have to fit between $0$ and $\lambda_\mathrm{max}$, and accordingly it increases polynomially with the number of oscillators $N$. 

The kinds of problems for which we can expect a polylogarithmic scaling in $N$ of the total query complexity necessarily need to satisfy $\lambda_{\mathrm{max}}/\Delta_{\lambda}^{(u)} =\mathcal{O}(\mathrm{polylog} (N))$ in our rigorous worst case scaling analysis.
On the one hand, the random glued-trees problem we discuss in \cref{sec:glued_tree} has this property, which ultimately originates from its very large degeneracy. 
This shows that there exists well-defined problems for which a large exponential speedup is possible. 
However, the task there is not the determination of response functions. 
On the other hand, if we simply consider a $\mathrm{1D}$ periodic chain with $N$ oscillators and unit spring constants and masses, the eigenvalues can be obtained analytically and are given by \cite{christandl2004}
\begin{equation*}
    \lambda_j 
    = 
    2 \left(
        1 
        - \cos\left(
            \frac{2 \pi (j-1)}{N}
        \right) 
    \right),
\end{equation*}
for $j \in \{1, \dots, N \}$. 
For this problem $s \norm{H}_{\mathrm{max}}$ is bounded, namely $s \norm{H}_{\mathrm{max}}=6$, but for large $N$ the gap for this problem scales as $\Delta_{\lambda} = \mathcal{O}(N^{-2})$, yielding eventually a polynomial scaling of the number of queries with $N$. 
Notice that these considerations have no influence on the scaling with the tolerance on the weights $\delta$, which is $1/\delta^3$ if we take it as an independent parameter (see \cref{eq:nquerytotscaling}).

\subsubsection{Possible improvements}
Possible improvements in the scaling in the $\delta$ parameter could be achieved combining our algorithm with the quantum amplitude estimation subroutine \cite{brassard2002}. Additionally, in the previous discussion, we have not restricted the domain of interest of the eigenvalues that are in general between $0$ and $\lambda_{\mathrm{max}}$. However, in practice we might be interested in evaluating the response function only in a certain frequency range for which only eigenvalues $\lambda_j \in [\lambda_a, \lambda_b]$ matter. In fact, we might know in advance the typical frequency spectrum of the forcing term, as it happens in the milling case discussed in \cref{subsec:motivation}. In this case, we can approximate the response function as
\begin{equation}
\label{eq:guufilter}
    G_{uu}(\lap) \approx \frac{1}{m_u}\sum_{j \in \mathcal{J}}\frac{W_{u j}^2}{\lambda_j + \lap^2},
\end{equation}
for $\mathrm{Im}(\lap) \in [\lambda_a, \lambda_b]$ and $\mathcal{J}
    =
    \{
        j:
        \lambda_j
        \in
        [
            \lambda_a,\lambda_b
        ]
    \}$. Thus, it might be beneficial to apply a filter to the quantum state $\ket{u}$ to remove all the contributions from eigenvalues that are outside the interval of interest $[\lambda_a, \lambda_b]$, and produce an initial, filtered state $\ket{u_f}$ such that
\begin{equation}
    \ket{u_f} \sim \sum_{\lambda_j \in [\lambda_a, \lambda_b]} W_{u j} \ket{\lambda_j}. 
\end{equation}
Quantum eigenvalue filtering has been already studied in the literature \cite{linlin2020, martyn2021, parrish2019} and could find an application in our problem. We note that also for classical algorithms it is possible to restrict the eigenvalue search in a specific range using filter diagonalization methods \cite{wall1995, chen1996, mandelshtam1997, IKEGAMI20101927}. 

Needless to say, our algorithm relies on the efficient implementation of the matrix oracles.
As discussed in \cref{subsec:complexity}, this usually requires the implementation of reversible, coherent arithmetic that comes at the cost of additional ancilla qubits. 
Alternatively, one could adapt state preparation techniques that avoid the need of coherent arithmetic \cite{sanders2019, macardle2022} to this task \footnote{In fact, the oracles are ultimately used to prepare the state in \cref{eq:psiu}}. 
In case the matrix elements are already stored in a classical memory, these data would need to be accessed by the oracle using a QRAM \cite{giovannetti2008}, which can lead to severe overheads destroying any exponential speedup (see discussion in \cite{dalzell2023, Aaronson2015}). 
For the milling use case discussed in \cref{subsec:motivation}, this would mean that the discretization of the problem would either need to happen in the quantum computer or it must be given in a format that the quantum computer can access efficiently. 
Rather than an intrinsic limitation, we believe that this is an interesting scientific question and underexplored area of research. 

Finally, we point out that our algorithm has some connection with the vector fitting algorithm \cite{gustavsen1999}, where the goal is, given $K$ observations of $G_{uu}(\lap)$ at $\lap =i \omega^{(k)}$, $k \in \{1, \dots, K \}$, to output the best estimate of the weights $W_{uj}^2$, usually called residues, and the eigenvalues $\lambda_j$. In our case, we want to achieve the same task, but using the data obtained from running QPE multiple times to estimate the weights and the eigenvalues.

\section{Quantum algorithm}
\label{sec:algo}
In this section, we detail how the local response function $G_{uu}(\lap)$ in ~\cref{eq:diagresponse} can be obtained using QPE. 
In particular, we describe the algorithm, and study its scaling, using the standard QPE \cite{NielsenChuang, cleve1998}, that makes use of the inverse quantum Fourier transform. 
A similar analysis can also be performed using other versions of QPE that use a single ancilla qubit for the phase register \cite{kitaev1995, kitaevBook, svore2014, martyn2021} or coherent QPE with fewer qubits \cite{rall2021}. 
As we have seen in \cref{subsec:problstate}, to obtain the response function, we need to solve the eigenproblem associated with the Hermitian matrix in \cref{eq:hamiltonian}. 

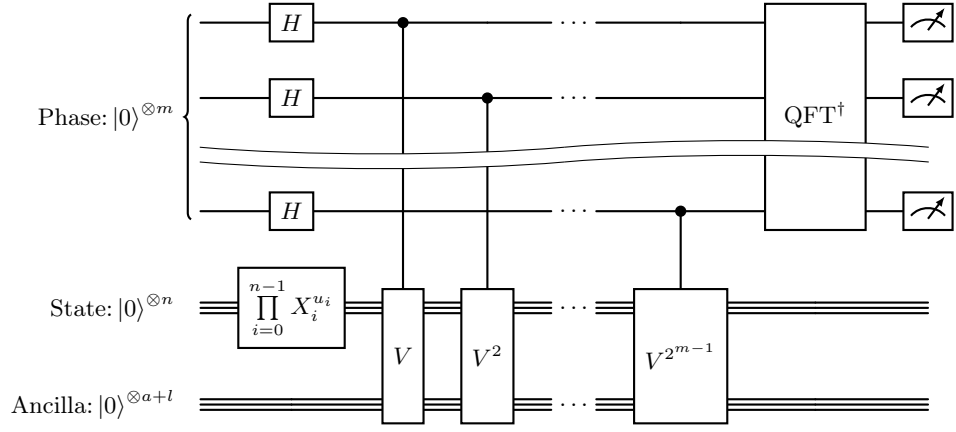
\begin{figure*}
	\centering
	\begin{adjustbox}{max width=0.99\linewidth}
		\begin{quantikz}[wire types= {q,q,q,q,b,b},classical gap=0.07cm]
			\lstick[wires=4, label style={anchor= east, xshift=-0cm}]{$\text{Phase:}\ket{0}^{\otimes m}$}& \gate{H}&
            \ctrl{5}&& \ \ldots\ && \gate[4,disable auto height]{\begin{array}{c}
					\\
					\text{$\text{QFT}^\dagger$} \\
					\\
			\end{array}}&  \meter{} 
			\\
			& \gate{H}&& \ctrl{4}& \ \ldots\ &&&\meter{}
			\\
			&\wave&&&&&&&
			\\
			& \gate{H}&&&\ \ldots\ & \ctrl{2}&&   \meter{}
			\\
			\lstick[label style={anchor= east, xshift = -0.2cm }]{$\text{State:}\ket{0}^{\otimes n}$}& \gate{\prod\limits_{i=0}^{n-1}X_i^{u_i}}
            & \gate[2, disable auto height]{V}&\gate[2, disable auto height]{V^2}&\ \ldots\ & \gate[2, disable auto height]{V^{2^{m-1}}}& &
			\\
			\lstick[label style={anchor= east, xshift=-0.2cm}]{$\text{Ancilla:}\ket{0}^{\otimes a+l}$}&&&& \ \ldots\ &&& 
		\end{quantikz}
	\end{adjustbox}
	\caption{QPE circuit for the computation of the local response function.
    The state preparation is realized by a series of NOT gates. QPE is performed using the walk operator $V$ defined in \cref{eq:qw_op} followed by the inverse quantum Fourier transform ($\mathrm{QFT}^{\dagger}$) \cite{NielsenChuang}.
    Sampling from the phase register and repeating many times allows us to estimate the relevant eigenvalues $\lambda_j$ and the weights $W_{uj}^2$ in \cref{eq:diagresponse}.
    }
	\label{fig:full_phase_estimation}
\end{figure*}

    \subsection{Matrix encoding}
    \label{subsec:hamiltonsim}
The algorithm is based on the use of QPE as eigenvalue estimator~\cite{abrams1999}. 
In our analysis, we focus on a version of QPE inspired by quantum walks~\cite{childs2010, berry2012} and similar to the one of Refs.~\cite{poulin2018, berry2018}. 
The basic idea of this method is that if we want to estimate the eigenvalues $\lambda$ of a Hamiltonian $H$, we could as well estimate a function $f(\lambda)$ as long as the function is invertible. 
In particular, in our case instead of performing QPE with unitary $e^{i H}$, we perform it using a unitary $V$ with eigenvalues $e^{\pm i \arccos (\alpha \lambda)}$ with $\alpha$ a normalization factor. 
The appeal of this method is that $V$ can be implemented exactly using the standard oracles, which simplifies the error analysis. 
The full QPE is shown in \cref{fig:full_phase_estimation}. 
In what follows, we briefly describe the method in the language of block-encoding and qubitization~\cite{Low2019hamiltonian, gilyen2019, martyn2021, tang2023, linlinnotes, dalzell2023}. 


Let us assume that we have a unitary $U_H$ acting on $a+n$ qubits that  realizes an $(\alpha, a, 0)$-block-encoding of our $n$-qubit Hamiltonian $H$. Mathematically, this translates into the requirement 
\begin{equation}
	\bra{0}^{\otimes a}U_H\ket{0}^{\otimes a} = \alpha H,
	\label{eq:block_encod}
\end{equation}
where $a$ is the number of additional ancilla qubits used for the block-encoding.  The eigenstates of $H$ with eigenvalue $\lambda$ will be denoted by $\ket{\lambda}$ in braket notation. Without loss of generality, we assume that $U_H$ is Hermitian, i.e., $U_H = U_H^{\dagger}$, which implies $U_H^2 = I_{a +n}$.\footnote{If $U_H$ is not Hermitian, one can add one ancilla qubit and define the Hermitian unitary $U_{H}^{(\mathrm{new})} = \ketbra{+}{-} \otimes U_H + \ketbra{-}{+}\otimes U_{H}^{\dagger}$, which is a $(\alpha, a+1, 0)$-block-encoding of $H$.}
We define the operator $\Pi$ as the projector onto the subspace with all ancillas in the zero state
\begin{equation}
    \Pi = \ketbra{0}{0}^{\otimes a} \otimes I_n,
\end{equation}
where $I_n$ denotes the $n$-qubit identity. 

The crucial property that qubitization uses is that the action of a Hermitian block-encoding $U_H$ on the state $\ket{0}^{\otimes a} \ket{\lambda}$ can be written as (see Section 10.4 in Ref.~\cite{dalzell2023})
\begin{equation}
    \label{eq:uhtolambda}
     U_H \ket{0}^{\otimes a}  \ket{\lambda} = \alpha \lambda \ket{0}^{\otimes a}  \ket{\lambda} + \sqrt{1 - \alpha^2 \lambda^2} \ket{0 \lambda^{\perp}},
\end{equation}
where $\ket{0 \lambda^{\perp}}$ is a state such that $(\bra{0}^{\otimes a}\bra{\psi})\ket{0 \lambda^{\perp}} =0$ $\forall$ $n$-qubit state $\ket{\psi}$ and $\braket{0 \lambda'^{\perp}}{0 \lambda^{\perp}} = \delta_{\lambda \lambda'}$. Additionally, 
\begin{equation}
\label{eq:uhtolambdaperp}
U_{H} \ket{0 \lambda^{\perp}} = \sqrt{1 - \alpha^2 \lambda^2} \ket{0}^{\otimes a}  \ket{\lambda} - \alpha \lambda \ket{0 \lambda^{\perp}}. 
\end{equation}
\cref{eq:uhtolambda} and \cref{eq:uhtolambdaperp} imply that $\forall \lambda$ we can identify a qubit-like subspace $\mathcal{H}^{(\lambda)} = \mathrm{span} \{\ket{0}^{\otimes a} \ket{\lambda}, \ket{0 \lambda^{\perp}}\}$, so that the action of $U_{H}$ onto this subspace can be represented by a $2 \times 2$ matrix $U_{H}^{(\lambda)}$ given by
\begin{equation}
\label{eq:ulambda}
   U_{H}^{(\lambda)} = \begin{pmatrix}
       \alpha \lambda & \sqrt{1 - \alpha^2 \lambda^2} \\
       \sqrt{1 - \alpha^2 \lambda^2} & -\alpha \lambda
   \end{pmatrix}. 
\end{equation}
However, the eigenvalues of this matrix are $\pm 1$, since $U_H$ is a reflection and thus, they do not carry information about the eigenvalues $\lambda$ of $H$. To remedy this issue we define the ``walk" operator $V$ that we use in the QPE subroutine depicted in \cref{fig:full_phase_estimation} as
\begin{equation}
    V =  U_H (2 \Pi - I_{a + n}).
    \label{eq:qw_op}
\end{equation}
In fact, the operator $2 \Pi - I_{a + n}$ also preserves the subspaces $\mathcal{H}^{(\lambda)}$ and its action on $\mathcal{H}^{(\lambda)}$ is simply represented by a Pauli $Z$ matrix:
\begin{equation}
    (2 \Pi - I_{a + n})^{(\lambda)} = \begin{pmatrix}
        1 & 0 \\
        0 & -1
    \end{pmatrix}.
    \label{eq:zero_refl}
\end{equation}
Consequently, the action of $V$ on $\mathcal{H}^{(\lambda)}$ is given by:
\begin{multline}
    V^{(\lambda)} = U_{H}^{(\lambda)} (2 \Pi - I_{a+n})^{(\lambda)}
    \\
    =
    \begin{pmatrix}
        \alpha \lambda & - \sqrt{1 - \alpha^2 \lambda^2} \\
        \sqrt{1 - \alpha^2 \lambda^2} & \alpha \lambda
    \end{pmatrix}.
\end{multline}
The matrix $V^{(\lambda)}$ has eigenvalues
\begin{equation}
    \mu_{\pm} = \alpha \lambda \pm i \sqrt{1- \alpha^2 \lambda^2} = e^{\pm i \arccos(\alpha \lambda)},
\end{equation}
with corresponding eigenvectors 
\begin{equation}
    \ket{\mu_{\pm}} = \frac{1}{\sqrt{2}} (\ket{0}^{\otimes a} \ket{\lambda} \mp i \ket{0 \lambda^{\perp}}).
    \label{eq:new_base}
\end{equation}
Thus, if we start from a state $\ket{\psi} = \sum_{\lambda} c_{\lambda} \ket{0}^{\otimes a}\ket{\lambda}$ and perform ``ideal" phase estimation using the operator $V$, the system will collapse in either $\ket{\mu_{+}}$ or $\ket{\mu_{-}}$ and the value $+\arccos(\alpha \lambda)$ or $-\arccos(\alpha \lambda)$, respectively, can be read out in the phase register. 
In either case, a subsequent measurement of the ancilla register in the computational basis would give all zeros with probability $1/2$ and thus, prepare the eigenstate $\ket{0}^{\otimes a} \ket{\lambda}$. 

The procedure we have just described works for an arbitrary Hermitian block-encoding of $U_H$. 
In \cref{app:quantumsubroutines}, we show for completeness how such a block-encoding can be realized if we have access to the oracles in \cref{eq:O_P} and \cref{eq:angle_oracle}. 
In this case, $a=n+2$  and the parameter $\alpha$ is given by
\begin{equation}
\label{eq:alpha}
    \alpha = \frac{1}{s \lVert H \rVert_{\mathrm{max}}},
\end{equation}
where $s$ denotes the sparsity of the matrix $H$.

    \subsection{Obtaining the response function}
    \label{subsec:algorithm}
In this section, we describe the full quantum algorithm to obtain the diagonal response function in \cref{eq:diagresponse}. As we see from \cref{fig:full_phase_estimation}, we have three main registers:
\begin{enumerate}
    \item An $m$-qubit register that, at the end of the circuit, stores the information about the eigenphase;
    \item An $n$-qubit register that represents the Hilbert space associated with the Hamiltonian $H$;
    \item An $(a+l)$-qubit ancilla register, which includes the $a$ ancillas for the block-encoding described in ~\cref{subsec:hamiltonsim} as well as $l$ ancilla qubits needed for the implementation of the oracles (see \cref{app:quantumsubroutines}). 
\end{enumerate}

We start by preparing a state $\ket{u}$ associated with one of the vertices $u\in \mathcal{V}$ in the state register (cf. \cref{fig:full_phase_estimation}) by using a product of single-qubit Pauli-$X$ gates
\begin{equation}
\label{eq:in_prod_state}
    \prod_{i=0}^{n-1} X^{u_i}_i \ket{0}^{\otimes n}
    =
    \ket{u}
    =
    \sum_{j=1}^{N}
    W_{uj} \ket{\lambda_j},
\end{equation}
where $X_i$ denotes the Pauli-$X$ gate on qubit $i$. 
This is equivalent to an amplitude encoding of the weights $W_{uj}$ in the basis of the eigenstates $\ket{\lambda_j}$ of $H$.
The choice of $u$ affects the set of weights $W_{uj}$ and with it the probability to collapse into the corresponding eigenstate $\ket{\lambda_j}$ upon measurement in this basis.

Including the $a$ ancilla qubits needed for the block-encoding, we can rewrite the state in \cref{eq:in_prod_state} in the eigenbasis of the  unitary $V$ given in \cref{eq:new_base} as
\begin{equation}
    \ket{0}^{\otimes a}\ket{u}
    =
    \frac{1}{\sqrt{2}}\sum_{j=1}^{N}
    W_{uj} \left(
         \ket{\mu_+^{(j)}}
        +\ket{\mu_-^{(j)}}
    \right).
\end{equation}
After the execution of QPE with $V$ as unitary operator controlled by the $m$ qubits of the phase register~\cite{NielsenChuang}, we get:
\begin{multline}
    \xrightarrow{\text{QPE}}
    \frac{1}{\sqrt{2}}\sum_{j=1}^{N} W_{uj} 
    \sum_{x=0}^{M-1} \left(
        a_{x+}^{(j)}\ket{\mu_+^{(j)}}
    \right.
    \\
    \left.
	   +a_{x-}^{(j)}\ket{\mu_-^{(j)}}
    \right)\ket{x},
    \label{eq:fin_state}
\end{multline}
where $M=2^m$ and $a_{x\pm}^{(j)}$ is the amplitude applied by the QPE subroutine
\begin{equation}
	a_{x \pm}^{(j)}
	=\frac{1}{M}
	\sum_{z=0}^{M-1}
	e^{
        \frac{2\pi i z}{M} 
        \left(
            \pm \varphi_j - x
	    \right)
    },
    \label{eq:qpe_amp}
\end{equation}
with
\begin{equation}
    \varphi_j
    = 
    \frac{M}{2\pi} \arccos \left( \frac{\lambda_j}{s \norm{H}_\text{max}} \right).
    \label{eq:phi_j}
\end{equation}
The coefficients $a_{x \pm}^{(j)}$ are peaked around $x= \pm \varphi_j \mod M$, respectively.
This tells us that the probability $P(x)$ to measure $x$ in the phase register will peak around $\pm \varphi_j \mod M$ for all $j\in\{1,\dots,N\}$, and it is given by
\begin{equation}
    P(x)
    =
    \sum_{j=1}^{N} \frac{W_{uj}^2}{2} 
    \left(
          \abs{a_{x+}^{(j)}}^2
    	   +\abs{a_{x-}^{(j)}}^2
    \right),
    \label{eq:prob_local}
\end{equation}
where we used the orthonormality of $|\mu_\pm^{(j)}\rangle$.

After sampling multiple times from the phase register, we can obtain an estimate of the probabilities~\cref{eq:prob_local}, from which we need to extract the eigenvalues $\lambda_j$ and the weights $W_{uj}^2$. 
Here, we assume that we have sufficient resolution to determine the peaks and their centers. 
Possible practical methods to determine the eigenvalues are discussed in Refs.~\cite{obrien_2019, svore2014, wiebe2016}, and could be tested in practice for our problem.   
The center $\tilde{\varphi}_j$ of the peak can be translated into the eigenvalues by inverting \cref{eq:phi_j}.
The probability to sample from a certain eigenstate is proportional to the weight $W_{uj}^2$ of interest.
The last can be estimated by adding the probabilities of the $2Q$ values closest to $\tilde{\varphi}_j$
\begin{equation}
    W_{uj}^2
    \approx
    2\sum_{q=-Q}^{Q-1}
        P(\lceil \varphi_j\rceil+q),
    \label{eq:weight_apprx}
\end{equation}
where, we assumed that $\tilde{\varphi}_j\in\left[\lfloor \varphi_j\rfloor,\lceil \varphi_j\rceil\right]$.
The additional factor of 2 is due to the $1/2$ in \cref{eq:prob_local}.
The necessary size of $Q$ is further discussed in \cref{subapp:weight_aprx}.
The full routine is gathered in Algorithm~\ref{alg:resp_func}.
\begin{algorithm}[t]
    \DontPrintSemicolon
    \caption{
        Calculation of the local response function $G_{uu}(\lap)$ on a quantum computer.
        The first part estimates classically the register and sample size required by the quantum circuit following in the loop over $t$.
        The classical post-processing in the end is required to extract the response function from the measurements.
    }
    \label{alg:resp_func}
    \KwData{
        \begin{itemize}
            \item Oscillator $u$
            \item Mass $m_u$
            \item Renormalization $s\norm{H}_\mathrm{max}$
            \item Minimum eigenvalue gap $\Delta_\lambda^{(u)}$
            \item Tolerances $\varepsilon$, $\delta$, $\zeta$
            \item Controlled quantum walk operator $cV$ which contains the oracles $O_\vartheta$ and $O_P$ 
            \item Number of oscillators $N$
            \item Number of distinguishable eigenvalues $N_u$
        \end{itemize}
    }
    \KwResult{Local response function $G_{uu}(\lap)$}
    \Begin{
        $m$, $n$, $a+l \gets$ compute the required register sizes\;
        $N_\mathrm{S} \gets$ compute the required sample size \;
        \For{$t \in \{1,\dots,N_\mathrm{S}\}$}{
            $\texttt{reg}_\mathrm{phase} \gets \sum_{i=0}^{2^m-1}\ket{i}$\;
            $\texttt{reg}_\mathrm{state} \gets \ket{u}$\;
            $\texttt{reg}_\mathrm{ancilla} \gets \ket{0}^{\otimes a+l}$\;
            \For{$i\in \{0,\dots,m-1\}$}{
                $
                cV^{2^i}$ to $
                    \texttt{[reg}_\mathrm{phase}^{(i)}, 
                    \texttt{reg}_\mathrm{state}, 
                    \texttt{reg}_\mathrm{ancilla}
                \texttt{]}
                $\;
            }
                            $ \mathrm{QFT}^\dagger $ to $\texttt{reg}_\mathrm{phase}$\;
            $\varphi_t \gets $ measure $\texttt{reg}_\mathrm{phase}$\;
        }
        define $P(\varphi) = \sum_{t=1}^{N_\mathrm{S}} \delta(\varphi-\varphi_t)/N_\mathrm{S}$\;
        $\bm{\varphi} \gets $ identify $N_u$ peaks in $P(\varphi)$\;
        \For{$j \in \{j:\varphi_j\in\bm{\varphi}\}$}{
            $\lambda_j\gets s\norm{H}_\mathrm{max}\cos(2^{1-m}\pi \varphi_j)$\;
            $W_{uj}^2\gets 2\sum_{q=-Q}^{Q-1} P(\lceil \varphi_j \rceil + q)$\;
        }
        $G_{uu}(\lap) \gets m_u^{-1}\sum_{j=1}^{N_u} W_{uj}^2/(\lambda_j +\lap^2) $\;
    }
\end{algorithm}

In \cref{app:nonlocal_resp}, we discuss a more general method which uses a modified Hadamard test to estimate the nonlocal response function $G_{uv}(\lap)$.


\section{Solving the random glued-trees problem}
\label{sec:glued_tree}
In this section, we show that a straightforward adaptation of the algorithm presented in \cref{sec:algo} can be used to solve the random glued-trees problem first studied in Ref.~\cite{childs2003} in the context of quantum walks. 
This problem is a well-known example for which quantum algorithms give a provable, oracular exponential speedup compared to classical methods. 
Recently, Ref.~\cite{babbush2023} also formulated the problem in terms of coupled harmonic oscillators showing that it can be solved in time polylogarithmic in the problem size, by using a particular encoding and Hamiltonian simulation. 
We here complement this result by showing that the problem can also be solved in polylogarithmic time using QPE. 
In what follows, we describe the problem and give our main results, while we provide the details in \cref{app:gluedtree}. 

\begin{figure}
\centering
\includegraphics[height=4cm]{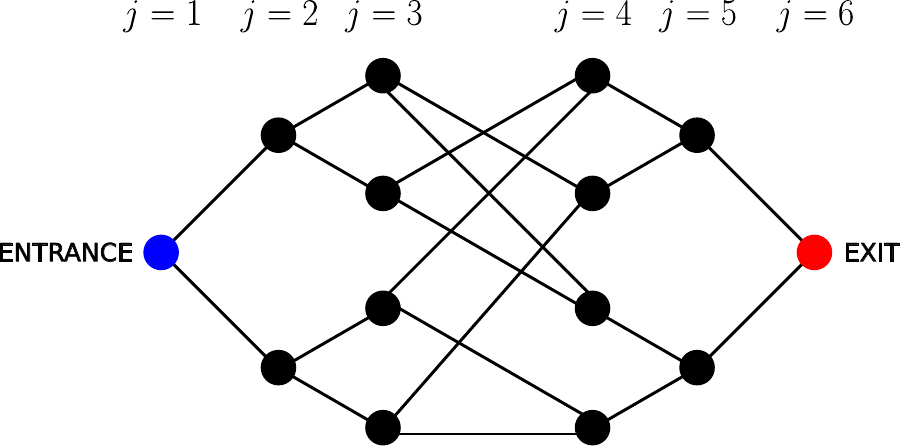}
\caption{Example of random glued-trees with $n_c=3$.}
\label{fig:randomglued}
\end{figure}

The basic setup of the random glued-trees problem is shown in \cref{fig:randomglued}. It consists of a graph $\mathcal{G} = (\mathcal{V}, \mathcal{E})$ with two balanced trees each with $n_c$ columns, where the $n_c$-th column of the right tree is glued randomly to  the $n_c$-th column of the left tree. In this way, each vertex in the $n_c$-th column of the right tree is connected to exactly two vertices of the $n_c$-th column of the left tree, and vice versa. Let us denote by $A$ the adjacency matrix of a random glued-trees graph. As we see in Fig.~\ref{fig:randomglued}, we denoted the root of the left tree as the ENTRANCE vertex, and the root of the right tree as the EXIT vertex. Access to the adjacency matrix $A$ allows us to identify the ENTRANCE and EXIT vertices, since they are the only two vertices with two neighbors. This allows us to define the following random glued-trees problem.

\begin{prob}
\label{prob:gluedtreeprob}
(Random glued-trees problem) Consider a random glued-trees graph with $N = 2 (2^{n_c} -1)$ vertices and $n_c$ the number of columns in each tree. Let $A$ denote its adjacency matrix. 
Let $\mathrm{deg}(i)$ be the function that given the row $i$ returns the degree of the associated vertex, i.e., the number of vertices connected to it. We assume that the vertices are randomly ordered, but we are given the index of the ENTRANCE vertex, i.e., the row associated with it. Assuming oracle access to the adjacency matrix $A$ and to the function $\mathrm{deg}(i)$ the goal is to find the index of the EXIT vertex.
\end{prob}

Note that the function $\mathrm{deg}(i)$ is merely used to identify the ENTRANCE and EXIT vertices, since they are the only vertices of degree $2$. \cref{prob:gluedtreeprob} has an immediate mapping to a system of coupled oscillators as discussed in Ref.~\cite{babbush2023}. In fact, we can associate with each vertex $u \in \mathcal{V}$ a unit mass so that the mass matrix is simply the identity matrix  $I_{N}$. Moreover, we add a unit spring constant to each edge in the graph and an additional unit spring constant to the ENTRANCE and EXIT masses attached to a wall. In this way, according to \cref{eq:hamiltonian}, the Hamiltonian $H$ coincides with the stiffness matrix and is given by $H=3 I_{N} - A$. Thus, the eigenvalues and eigenvectors of $H$ are in one-to-one correspondence with those of the adjacency matrix $A$. In what follows, we assume that we have a block-encoding of $A$, which is $3$-sparse and has $\lVert A \rVert_{\mathrm{max}} =1$.

The mapping of the problem to a quantum computer is the same as the one described in \cref{subsec:mainres} for any coupled oscillator problem.
To solve it using QPE, we start by preparing the computational basis string associated with the ENTRANCE vertex (known to us) that we denote as $\ket{\mathrm{ENTRANCE}}$. 
The idea is now to perform QPE as in \cref{fig:full_phase_estimation}, using a block-encoding of the adjacency matrix $A$. 
The glued-trees problem enjoys an effective, exponential dimensionality reduction and under the action of the adjacency matrix the evolution of the system is confined to the \emph{column subspace} $\mathcal{H}_{\mathrm{col}}$ (see \cref{app:gluedtree} for more details). 
Following Ref.~\cite{childs2003}, one can obtain an estimate for the minimum eigenvalue gap $\Delta_{\lambda}$ of the adjacency matrix projected onto the column subspace $A_{\mathrm{col}}$ defined in \cref{eq:api} \footnote{Ref.~\cite{childs2003} gives the result in a slightly different form, namely stating that the eigenvalue spacing is larger than $\frac{2 \pi^2}{(1 + \sqrt{2})n_c^3} + \mathcal{O}\left(\frac{1}{n_c^4}\right)$. 
However, adapting the derivation one can show that \cref{eq:gluedgap} also holds.}:
\begin{equation}
\label{eq:gluedgap}
    \Delta_{\lambda} = \frac{8 \pi^2}{n_c^3} + \mathcal{O}\left(\frac{1}{n_c^4} \right).
\end{equation}
\cref{eq:gluedgap} holds for any random glued-trees graph. Thus, we see that, despite the fact that the size of the adjacency matrix increases exponentially with the number of columns $n_c$, i.e., with the number of qubits $n$ needed to encode it, the relevant, minimum gap scales only inverse polynomially. This is exactly one of the features needed for a possible exponential quantum advantage. The idea now is to take the number of qubits $m$ in the phase register large enough to resolve the minimum eigenvalue gap $\Delta_{\lambda}$. 
From \cref{eq:m_min_eval}, we require
\begin{equation}
\label{eq:mglued}
    m = \gamma \left\lceil \log_2 \left(\frac{3 \pi}{\Delta_{\lambda}} \right) \right \rceil \approx \gamma \left\lceil \log_2  \left( \frac{3 n_c^3}{8\pi} \right)\right \rceil,
\end{equation}
with $\gamma$ a constant, integer factor of our choice such that $\gamma \gg 1$. 
Accordingly, the number of oracle queries for a single run of QPE will be of order
\begin{equation}
    N_{\mathrm{queries}} = \mathcal{O}(n_c^3).
\end{equation}

If \cref{eq:mglued} is satisfied, then the QPE subroutine described in \cref{sec:algo}, followed by measurement of the ancilla register with post-selection on the $\ket{0}^{\otimes a}$ outcome, behaves de facto as a measurement in the eigenbasis of $A$ projected onto the column subspace. Thus, after one run of QPE followed by post-selection on $\ket{0}^{\otimes a}$, we have prepared an eigenstate of the adjacency matrix $A$ projected onto the column subspace in the $n$-qubit register. 
A final measurement in the computational basis on this register gives a vertex $u \in \mathcal{V}$. 
We are interested in the probability $\mathrm{Prob}(\mathrm{EXIT})$ that this procedure outputs the EXIT vertex. 
In \cref{app:gluedtree}, we obtain an analytical lower bound for this probability (see \cref{eq:lowerboundpexit}), from which we conclude that 
\begin{equation}
\label{eq:gluedscaling}
    \mathrm{Prob}(\mathrm{EXIT}) = \mathcal{O} \left( \frac{1}{n_c}\right),
\end{equation}
which finally proves the overall polynomial scaling of the algorithm in the number of columns $n_c$, i.e., polylogarithmic in the size of $A$, complementing the results obtained in  Refs.~\cite{babbush2023,childs2003} with time-evolution-based algorithms.

\section{Conclusions}
\label{sec:conclusion}

We have presented a quantum algorithm for obtaining response functions on systems of coupled harmonic oscillators. 
These kinds of problems also effectively emerge in the study of mechanical, vibrating structures, after discretization, as well as other areas of interest. 
The particular analytical structure of these functions makes them suitable for a quantum computer. The algorithm is based on the QPE subroutine, that we use as an eigenvalue solver. 
Importantly, the state preparation at the input of the algorithm just amounts to the preparation of a product state. 
This allows us to analyze the scaling of the algorithm in terms of space and queries of the matrix oracles. 
While the algorithm achieves an exponential saving in space, the scaling of the total number of oracles queries is problem-dependent and can be polynomial in the number of qubits only when very specific conditions are met. We identify such a list of sufficient conditions leading to a rigorous worst case guarantee, but also leave open the possibility for improved practical scaling when a priori information about the structure of the problem is available.
In addition, the scaling is mostly limited by statistical error coming from the estimation of the weights of the response functions. 
Possible improvements of the algorithm can be obtained by combining the algorithm with quantum eigenvalue filtering techniques, as well as quantum amplitude estimation. 
Finally, we have shown that the algorithm can be adapted to efficiently solve the random-glued trees problems, similarly to other approaches based on Hamiltonian simulation. 

While we performed a basic scaling analysis, we believe that a deeper understanding of our algorithm and its potential can be obtained by studying its performances in practice. 
This could be simply understood via basic classical simulations that aim at obtaining the minimum, relevant eigenvalue gap for some class of oscillator problems. 
Such analysis would also shed light on the kinds of problems for which large advantages are possible. Additionally, a practical study of the algorithm performances using methods developed for the estimation of multiple eigenvalues \cite{obrien_2019, svore2014, wiebe2016} tailored to our problem, would be beneficial. 
Finally, an open question is whether estimates of the response functions can also be obtained using time-evolution based algorithms for coupled harmonic oscillator as in Ref.~\cite{babbush2023}.

\section*{Acknowledgements}
SD, SS and PK are funded by the Bundesministerium für Wirtschaft und Klimaschutz (BMWK, Federal Ministry for Economic Affairs and Climate Action) in the quantum computing enhanced service ecosystem for simulation in manufacturing project (QUASIM, Grant No. 01MQ22001A).
AC acknowledges funding from the Deutsche Forschungsgemeinschaft (DFG, German Research Foundation) under Germany’s Excellence Strategy - Cluster of Excellence Matter and Light for Quantum Computing (ML4Q) EXC 2004/1 - 390534769. AC acknowledge funding from the German Federal Ministry of Education and Research (BMBF) in the funding program ``Quantum technologies - from basic research to market'' (Project QAI2, contract No. 13N15585).
MB is supported by the EPSRC Grant No. EP/W032643/1 and the Excellence Cluster - Matter and Light for Quantum Computing (ML4Q).

\section*{Data availability}
The algorithm presented in this manuscript were tested for the example of linearly coupled oscillators.
For this case, it was implemented and emulated from end-to-end in Qiskit~\cite{Qiskit}, including compilation of the matrix oracles.
The corresponding repository containing all the required packages and a demonstration notebook can be found at \href{https://go.fzj.de/resp-QPE}{https://go.fzj.de/resp-QPE}.

\bibliography{biblio_quafem}

\newpage

\appendix 
\crefalias{section}{appendix}

\numberwithin{equation}{section}

\section{Analytical form of the response functions}
\label{app:responseform}
In this appendix, we show that for a system of coupled harmonic oscillators the matrix elements of the response function $G_{uv}(\lap)$ in Laplace domain can be written as 
\begin{equation}
\label{eq:guv}
    G_{uv}(\lap) = \frac{1}{\sqrt{m_u m_v}}\sum_{j=1}^N \frac{
        W_{u j} W_{v j}
    }{
        \lambda_j 
        + \lap^2
    }.
\end{equation}
We remark that this is a well-known fact in the study of harmonic systems, in particular in the theory of electrical circuits, where it appears in various forms~\cite{newcomb1966linear, russer, foster1924}. Also, it is the base for the black-box quantization method for quantum electrical circuits~\cite{Nigg.etal.2012:BlackBoxCqed}. Here we provide a derivation for our problem adapted from Ref.~\cite{ciani2024}.

Let us consider the case in which we apply to oscillator $v$ an impulse $f_v(t) = \tilde{f}_v\delta(t)$, while no force is applied to the other oscillators, i.e.,  $f_{v'}(t)=0$ for $v' \neq v$. 
In terms of the normal mode variables in \cref{eq:yvar}, the dynamical equation in time domain \cref{eq:dyneqf} reads
\begin{equation}
    \frac{d^2 y_j}{dt^2} = - \lambda_j y_j + \frac{W_{vj}}{\sqrt{m_v}} \tilde{f}_v\delta(t), 
\end{equation}
which in Laplace domain gives
\begin{equation}
    Y_{j}(\lap) = \frac{1}{\sqrt{m_v}}\frac{W_{v j}}{ \lambda_j + \lap^2}\tilde{f}_v,
\end{equation}
for $j\in\{1, \dots, N\}$. From \cref{eq:xwujy}, we get that the Laplace transform of $x_u(t)$ can be written as
\begin{multline}
    X_u(\lap) = \frac{1}{\sqrt{m_u}} \sum_{j=1}^N W_{uj} Y_{j}(\lap) \\
    = \frac{1}{\sqrt{m_u m_v}} \sum_{j=1}^N \frac{W_{uj} W_{vj}}{\lambda_j+ \lap^2}\tilde{f}_v,
\end{multline}
from which \cref{eq:guv} follows. 

\section{Hermitian block-encoding from the matrix oracles}
\label{app:quantumsubroutines}
Here we detail how a Hermitian $(\alpha, a, 0)$-block-encoding $U_H$ of a Hamiltonian $H$ can be constructed using the matrix oracles $O_P$ and $O_{\vartheta}$ defined in \cref{eq:O_P} and \cref{eq:angle_oracle}, respectively. In particular, we limit ourselves to the case when $H$ is real, i.e., symmetric, and all its diagonal elements are positive, which is the case for the matrix in \cref{eq:hamiltonian} we are interested in. We start by considering a $2(n+1)$-qubit register, where the first $a=n+2$ qubits will be the ancilla qubits needed for the block-encoding. We take the block-encoding unitary $U_H$ to be of the following form
\begin{equation}
    U_H = U_T^{\dagger} \mathrm{SWAP}_{n+1} U_T,
    \label{eq:block_encode_USU}
\end{equation}
where $\mathrm{SWAP}_{n+1}$ swaps two $(n+1)$-qubit registers, i.e., $\mathrm{SWAP}_{n+1} \ket{\psi} \ket{\varphi} = \ket{\varphi} \ket{\psi}$ for any $(n+1)$-qubit states $\ket{\psi}, \ket{\varphi}$. Thus, Eq~\eqref{eq:block_encod} and Eq.~\eqref{eq:block_encode_USU} imply that
\begin{equation}
\label{eq:beswapcond}
    \alpha H = \bra{0}^{n+2} U_T^{\dagger} \mathrm{SWAP}_{n+1} U_T \ket{0}^{\otimes n+2}.
\end{equation}
We further assume that the map $U_T \ket{0}^{\otimes n+2}$, which maps an $n$-qubit quantum state to a $2(n+1)$-qubit quantum state can be written as ($N=2^n$)
\begin{equation}
\label{eq:ut1}
    U_T \ket{0}^{\otimes n+2} = \sum_{u=1}^{N} \ket{\psi_u, 0, u} \bra{u},
\end{equation}
where $\ket{\psi_u}$ denotes an $(n+1)$-qubit state to be determined, $\ket{0}$ the single-qubit zero state and $\ket{u}$ an $n$-qubit computational basis state. Plugging \cref{eq:ut1} into \cref{eq:beswapcond}, we obtain that the states $\ket{\psi_u}$ need to satisfy
\begin{equation}
    \alpha H_{u v} = \braket{\psi_u, 0, u}{0, v, \psi_v}.
\end{equation}
A possible solution is to set $\alpha$ as in \cref{eq:alpha} and take
\begin{multline}
\label{eq:psiu}
    \ket{\psi_u} 
    = 
    \frac{1}{\sqrt{s}} \sum_{v \in \tilde{N}_\mathcal{G}[u]} 
        (i \mathrm{sgn}(u-v))^{\Theta(-H_{uv})} 
        \\
        \times \left(
            \sqrt{\frac{|H_{uv}|}{\Vert H\rVert_{\mathrm{max}}}} \ket{0} 
            +\sqrt{1 - \frac{|H_{uv}|}{\Vert H\rVert_{\mathrm{max} }}}\ket{1}
        \right)\ket{v},
\end{multline}
where $\mathrm{sgn}(\cdot)$ is the sign function.\footnote{We set $\Theta(0)=0$ and $\mathrm{sgn}(0)=1$.}
Note that \cref{eq:psiu} requires the diagonal elements to be non-negative to give a valid block-encoding.
If this is not the case, one can always enforce this property by adding a suited matrix proportional to the identity to $H$.  

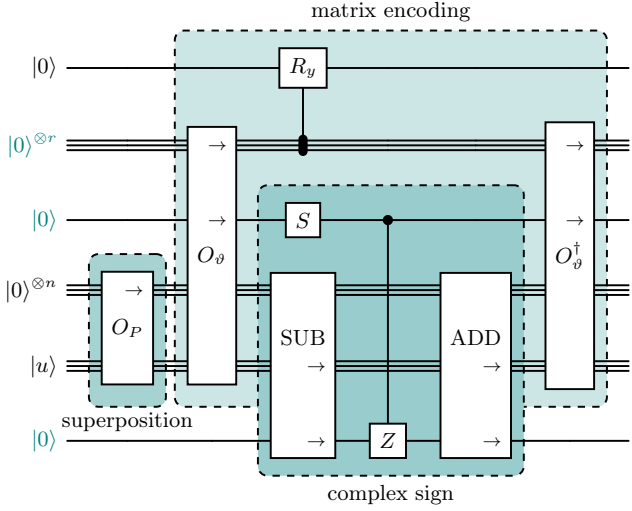
\begin{figure}
	\centering
	\adjustbox{max width = \linewidth}{
		\begin{quantikz}[ wire types={q,b,q,b,b,q}, classical gap =0.07cm]
			\lstick{$\ket{0}$}&& \gategroup[5,steps=5,style={dashed,rounded
corners,fill=teal!20, inner xsep=2pt}, background, label style={label position=above, anchor=south,yshift=-0.2cm}]{{matrix encoding}}&\gate{R_y}&&&&
			\\
			\lstick{\textcolor{teal}{$\ket{0}^{\otimes r}$}}&& \gate[4]{O_\vartheta}   \gateoutput{$\rightarrow$}& \ctrl{-1}&&& \gate[4]{O_\vartheta^\dagger}   \gateoutput{$\rightarrow$}&
			\\
			\lstick{\textcolor{teal}{$\ket{0}$}}&& \gateoutput{$\rightarrow$}& \gate{S}\gategroup[4,steps=3,style={dashed,rounded
corners,fill=teal!40, inner xsep=2pt}, background, label style={label position=below, anchor=north,yshift=-0.2cm}]{{complex sign}}&  \ctrl{3}&&\gateoutput{$\rightarrow$}&
			\\
			\lstick{$\ket{0}^{\otimes n}$}& \gate[2]{O_P} \gategroup[2,steps=1,style={dashed,rounded
corners,fill=teal!35, inner xsep=2pt}, background, label style={label position=below, anchor=north,yshift=-0.15cm}]{{superposition}}\gateoutput{$\rightarrow$} && \gate[3,label style={yshift=4mm}]{\text{SUB}}&& \gate[3,label style={yshift=4mm}]{\text{ADD}}&&
			\\
			\lstick{$\ket{u}$}&&&   \gateoutput{$\rightarrow$}&&  \gateoutput{$\rightarrow$}&&
            \\
			\lstick{\textcolor{teal}{$\ket{0}$}}&&&  \gateoutput{$\rightarrow$}& \gate{Z}&  \gateoutput{$\rightarrow$}&&
		\end{quantikz}
	}
	\caption{
        Circuit for the state transformation $U_T$.
        It consists of three parts.
        The preparation of the superposition with $O_P$, the matrix encoding in the amplitude and the multiplication with the complex sign.
        The order of the registers in here is chosen for a compact circuit layout.
        The first and fourth register combine into $\ket{\psi_u}$ of \cref{eq:ut1}.
        The fifth register is $\ket{u}$ in \cref{eq:ut1}.
        The second, third and last registers (in \textcolor{teal}{blue}) are of auxiliary nature and will be reinitialized after the full block-encoding of $H$.
        In the second register we store the intermediate angle (\cref{eq:anglehuv}).
        The third register is used for the binary sign $\Theta(-H_{uv})$ and the last for $\Theta(\mathrm{sgn}(u-v))$.
    }
	\label{fig:state_trafo}
\end{figure}

Given the form of the block-encoding in \cref{eq:block_encode_USU}, all we need to show is how to construct a unitary $U_T$ that satisfies \cref{eq:ut1} with states $\ket{\psi_u}$ given in \cref{eq:psiu}. The implementation of $U_T$ consists of three parts. First preparing a superposition of all states $\ket{v}$ with non-zero matrix entries $H_{uv}$. Second, encoding the matrix entries in the amplitudes,
and last, preparing the corresponding signs. All these steps are shown in \cref{fig:state_trafo} as a quantum circuit that involves the matrix oracles $O_P$ and $O_{\vartheta}$.   

We now describe in detail each transformation. 
As we see from \cref{fig:state_trafo} we have a total of six different registers. The three register highlighted in blue are additional ancilla registers needed for the implementation of the oracles, namely:
\begin{itemize}
    \item an $r$-qubit ancilla register to store the value of the angles $\vartheta_{uv}$ given in \cref{eq:anglehuv};
    \item an ancilla qubit to store the value of the sign of the matrix elements $H_{uv}$;
    \item an ancilla qubit (bottom one) to store the value of the sign of $u-v$.
\end{itemize}
The remaining registers form a $(2n+1)$-qubit register that together with an additional qubit (not shown in the figure) will be used to implement the $2(n+1)$-qubit block-encoding we described at the beginning of this appendix.

The system starts in the state $\ket{0}^{\otimes r +2 +n} \ket{u}\ket{0}$. After, the application of the position oracle $O_P$ defined in \cref{eq:O_P} the state becomes
\begin{equation}
    \xrightarrow{O_P}
     \frac{1}{\sqrt{s}}
     \sum_{v \in \tilde{N}_{\mathcal{G}}[u]}
        \ket{0}^{\otimes r+2} 
        \ket{v}\ket{u}\ket{0},
\end{equation}
and after the oracle $O_{\vartheta}$ given in \cref{eq:angle_oracle}
\begin{equation}
    \xrightarrow{O_{\vartheta}} \frac{1}{\sqrt{s}}\sum_{v \in \tilde{N}_{\mathcal{G}}[u]} \ket{0} \ket{\vartheta_{uv}, \Theta(-H_{uv})} \ket{v, u} \ket{0}.
\end{equation}

We continue by reading the angle $\vartheta_{uv}$ in the ancilla register with a series of controlled $y$-rotations $\mathrm{c}R_y=\prod_{k=1}^{r} \mathrm{c}^l R_y(2^{-k}\pi)$ with target the ancilla qubit on the top of \cref{fig:state_trafo} and control the $k$-th qubit in the binary representation of $\ket{\vartheta_{uv}}$ in \cref{eq:rbitthetaket}.\footnote{
    We use the standard definition for the $y$-rotation in here $R_y(\vartheta) = e^{-i\vartheta\sigma_y/2}$, where $\sigma_y$ is the second Pauli matrix.
}
This gives
\begin{multline}
    \xrightarrow{\mathrm{c}R_y}
    \frac{1}{\sqrt{s}}\sum_{v \in \tilde{N}_\mathcal{G}[u]}
    \left(
        \sqrt{
            \frac{
                \abs{H_{uv}}
            }{
                \norm{H}_\text{max}
            }
        }\ket{0}
        +\sqrt{
            1
            -\frac{
                \abs{H_{uv}}
            }{
                \norm{H}_\text{max}
            }
        }
        \ket{1}
    \right)
    \\
    \otimes
    \ket{\vartheta_{uv},\Theta(-H_{uv})}
    \ket{v,u} 
    \ket{0}.
\end{multline}

It remains to implement the coefficient $i\text{\,sgn}(u-v)$ for negative $H_{uv}$. The phase $i^{\Theta(-H_{uv})}$ can be realized with an S gate to the sign-qubit $\ket{\Theta(-H_{uv})}$:
\begin{multline}
    \xrightarrow{S}
    \frac{1}{\sqrt{s}}\sum_{v \in \tilde{N}_\mathcal{G}[u]} i^{\Theta(-H_{uv})}\\
    \times \left(
        \sqrt{
            \frac{
                \abs{H_{uv}}
            }{
                \norm{H}_\text{max}
            }
        }\ket{0}
        +\sqrt{
            1
            -\frac{
                \abs{H_{uv}}
            }{
                \norm{H}_\text{max}
            }
        }
        \ket{1}
    \right)
    \\
    \otimes
    \ket{\vartheta_{uv},\Theta(-H_{uv})}
    \ket{v,u} 
    \ket{0}.
\end{multline}
For the relative sign, we subtract the $\ket{v}$ register from the $\ket{u}$ for which we need to increase the target register by one qubit to store the sign
\begin{equation}
    \ket{v}\ket{u,0}
    \xrightarrow{\text{SUB}}
    \ket{v}\ket{u-v \mod 2^{n+1}}.
\end{equation}
Notice that the most significant bit of $u-v \mod 2^{n+1}$, which is a $(n+1)$-bit number, encodes the sign of the regular subtraction $u-v$, since both $u$ and $v$ are $n$-bit numbers. Thus, we can write 
\begin{equation}
    \ket{u-v \mod 2^{n+1}} = \ket{u-v \mod 2^n} \ket{\Theta(v-u)},
\end{equation}
where the register associated with $\ket{u-v \mod 2^n}$ is an $n$-qubit register.
The subtraction gives the state
\begin{multline}
    \xrightarrow{\mathrm{SUB}}
    \frac{1}{\sqrt{s}}\sum_{v \in \tilde{N}_\mathcal{G}[u]} i^{\Theta(-H_{uv})}\\
    \times \left(
        \sqrt{
            \frac{
                \abs{H_{uv}}
            }{
                \norm{H}_\text{max}
            }
        }\ket{0}
        +\sqrt{
            1
            -\frac{
                \abs{H_{uv}}
            }{
                \norm{H}_\text{max}
            }
        }
        \ket{1}
    \right)
    \\
    \otimes
    \ket{\vartheta_{uv},\Theta(-H_{uv})}
    \ket{v, u-v \mod 2^n}
    \ket{\Theta(v-u)}.
\end{multline}
A controlled $Z$ gate with control the qubit in the state $\ket{\Theta(-H_{uv})}$ and target the one in the state $\ket{\Theta(v-u)}$ (represented by the qubit at the bottom in \cref{fig:state_trafo}) yields the desired coefficient
\begin{multline}
    \xrightarrow{\mathrm{c}Z}
    \frac{1}{\sqrt{s}}\sum_{v \in \tilde{N}_\mathcal{G}[u]}  (i \mathrm{sgn}(u-v))^{\Theta(-H_{uv})} \\
    \times \left(
        \sqrt{
            \frac{
                \abs{H_{uv}}
            }{
                \norm{H}_\text{max}
            }
        }\ket{0}
        +\sqrt{
            1
            -\frac{
                \abs{H_{uv}}
            }{
                \norm{H}_\text{max}
            }
        }
        \ket{1}
    \right)
    \\
    \otimes
    \ket{\vartheta_{uv},\Theta(-H_{uv})}
    \ket{v, u-v \mod 2^n}
    \ket{\Theta(v-u)}.
 \end{multline}
At last we invert the subtraction SUB using the adder ADD
\begin{multline}
    \xrightarrow{\mathrm{ADD}}
    \frac{1}{\sqrt{s}}\sum_{v \in \tilde{N}_\mathcal{G}[u]}  (i \mathrm{sgn}(u-v))^{\Theta(-H_{uv})} \\
    \times \left(
        \sqrt{
            \frac{
                \abs{H_{uv}}
            }{
                \norm{H}_\text{max}
            }
        }\ket{0}
        +\sqrt{
            1
            -\frac{
                \abs{H_{uv}}
            }{
                \norm{H}_\text{max}
            }
        }
        \ket{1}
    \right)
    \\
    \otimes
    \ket{\vartheta_{uv},\Theta(-H_{uv})}
    \ket{v, u}
    \ket{0},
\end{multline}
and the oracle $O_\vartheta$ using $O_\vartheta^{\dagger}$
\begin{multline}
    \xrightarrow{O_\vartheta^\dagger}
    \frac{1}{\sqrt{s}}\sum_{v \in \tilde{N}_\mathcal{G}[u]}  (i \mathrm{sgn}(u-v))^{\Theta(-H_{uv})} \\
    \times \left(
        \sqrt{
            \frac{
                \abs{H_{uv}}
            }{
                \norm{H}_\text{max}
            }
        }\ket{0}
        +\sqrt{
            1
            -\frac{
                \abs{H_{uv}}
            }{
                \norm{H}_\text{max}
            }
        }
        \ket{1}
    \right)
    \\
    \otimes
    \ket{0}^{\otimes r+1}
    \ket{v, u}
    \ket{0},
\end{multline}
and thus, we have effectively prepared the state $\ket{\psi_u}$ in \cref{eq:psiu}. 
Notice that in this procedure the $r+1$ ancilla qubits needed for the implementation of $O_{\vartheta}$ and the ancilla qubit that stores the sign of $u-v$ return to the zero state, and this is also the case for the full block-encoding in \cref{eq:block_encod}. This is why we should not count them among the $a$ ancilla qubits of the block-encoding, which are thus $a=n+2$.

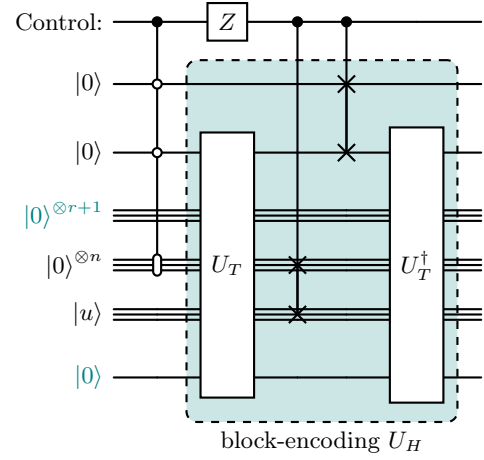
\begin{figure}
	\centering
	\adjustbox{max width = \linewidth}{
		\begin{quantikz}[wire types={q,q,q,b,b,b,q}, classical gap = 0.07cm]
			\lstick  {Control:}& \control{}& \gate{Z}& \ctrl{5}& \ctrl{2} &&
            \\
			\lstick  {$\ket{0}$}&\octrl{-1}&\gategroup[
                6,steps=4,
                style={dashed,rounded
corners,fill=teal!20, inner xsep=2pt}, 
                background, 
                label style={label position=below, anchor=north,
                yshift=-0.2cm}]{{block-encoding $U_H$}}&&\swap{1}&&
			\\
			\lstick  {$\ket{0}$}&  \octrl{-1}&\gate[5]{U_T}&&  \targX{}&\gate[5]{U_T^\dagger}&  
			\\
			\lstick{\textcolor{teal}{$\ket{0}^{\otimes r+1}$}}&&&&&&  
			\\
			\lstick{$\ket{0}^{\otimes n}$}& \octrl{-2}&&   \swap{1}&&&  
			\\
			\lstick  {$\ket{u}$}&&&\targX{}&&&
			\\
			\lstick{\textcolor{teal}{$\ket{0}$}}&&&&&&  
		\end{quantikz}
	}
	\caption{
        Controlled $V$ operator with $V$ defined in \cref{eq:qw_op}.
        The circuit starts with a controlled reflection around $\ket{0}^{\otimes n+2}$ that is implemented by a multi-controlled and a normal $Z$ gate to the control qubit.
        After this we see the controlled version of the block-encoding $U_H = U_T^\dagger\mathrm{SWAP}_{n+1}U_T$.
    }  
	\label{fig:circ_cqw}
\end{figure}

The circuit implementation of $U_H$ in the block-encoding \cref{eq:block_encode_USU} is shown in \cref{fig:circ_cqw}.
Note that the total number of qubits was increased by one.
This is the last one of the $a=n+2$ ancilla qubits necessary for the Hilbert-space extension.
The operator in front is reflecting around $\ket{0}^{\otimes n+2}$ and transforms $U_H$ into the quantum walk operator defined in \cref{eq:qw_op}.
It consists of a multi-controlled and a regular $Z$ gate.
The QPE demands a controlled version which is why we have another qubit which controls the CNOT gate in the reflection and the SWAP gates in $U_H$.

    \subsection{Qubit requirements}
    \label{subapp:qubit}

The full QPE estimation is applied to three registers.
The $m$-qubit phase register, the $n$-qubit state register and the $(a+l)$-ancilla register used by the controlled walk operator $\mathrm{c}V$.
We will focus on the $n+a+l+1$ qubits required for the implementation of $\mathrm{c}V$.
For this we start with $U_T$ described in \cref{fig:state_trafo}
.
The total number of qubits $n_{U_T}$ required for the state transformation is given by
\begin{equation}
\label{eq:nqubitut}
    n_{U_T}
    =
    2n+1+\max(a_P,r+1+a_\vartheta,r+2+a_\mathrm{ADD}),
\end{equation}
where $(r+1)$ is the number of qubits used by $O_\vartheta$ to store the angle $\vartheta_{uv}$ and the sign of $H_{uv}$.
Similar to $m$ it effects the accuracy of the results. Moreover,
the number of ancilla qubits necessary for the position and angle oracles, and the quantum adder are given by $a_P$, $a_\vartheta$ and $a_\mathrm{ADD}$, respectively. 
In \cref{eq:nqubitut}, we used that the ancilla qubits used in $O_P$ can be reused in $O_\vartheta$, while the ancilla qubits used there can be used again in $O_\mathrm{ADD}$.

The state transformation $U_T$ appears twice in the circuit for $\mathrm{c}V$ (cf. \cref{fig:circ_cqw}).
The circuit is further extended by a control qubit and the last qubit necessary for the block-encoding.
Hence, we have a total number of $n_{\mathrm{c}V}$ qubits necessary for the controlled quantum walk, which is
\begin{equation}
    n_{\mathrm{c}V}
    =
    2n+3+\max(a_P,r+1+a_\vartheta,r+2+a_\mathrm{ADD}, n).
\end{equation}
Here we see additional $n$ ancilla qubits, that are required for the Toffoli gates in the mutli-controlled Z gate in \cref{fig:circ_cqw}.
Those can be reused again.
In total we have $m$ repetitions of $cV$ with varying control qubit which leads to the scaling described in \cref{eq:ntotscaling}.
There we replaced $m$ with the scaling \cref{eq:mminmax} described in \cref{app:detail_error}.

\section{Detailed error analysis}
\label{app:detail_error}

In this appendix, we discuss the effect of discretization and statistical errors on the estimated eigenvalues $\tilde{\lambda}_j$ and weights $\tilde{W}_{uj}^2$ and how one can control the maximum error by increasing the size of the phase register. This analysis proves the statement of \cref{thm:mainthm}.

In particular, we analyze the effect of the discretization of the phase when it is stored as binary number in the $m$-qubit register with focus on the eigenvalue in \cref{subapp:err_phase} and the weights in \cref{subapp:weight_aprx}. The size of the phase register determines the number of oracle queries necessary to implement the $\mathrm{controlled}$-$V^{2^k}$ operations in Algorithm~\ref{alg:resp_func} shown in \cref{fig:full_phase_estimation}. We discuss this connection in \cref{subapp:oraclequeries}. We point out that the number of oracle queries constitutes the main limitation of the algorithm in most \emph{practical} cases (see discussion in \cref{subsec:algolimit}).
At last, in \cref{subapp:sample_size}, we describe the connection between the sample size and the statistical accuracy of the estimated weights.
In the following, for notational simplicity, we focus on the non-degenerate systems, but the results carry over to the degenerate case as discussed in \cref{subsec:mainres}.

    \subsection{Phase register size for given tolerances}
    \label{subapp:phase_register_size}
\subsubsection{Eigenvalue approximation}
\label{subapp:err_phase}
The following analysis is more easily understood by considering the case in which we perform QPE using the operator $V$ as in \cref{fig:full_phase_estimation}, but taking as initial state in the $n$-qubit register a single eigenstate of $H$ $\ket{\lambda}$. When we measure the $m$-qubit phase register at the end of QPE we get a bitstring that represents an integer number between $0$ and $M-1$ with $M=2^m$. This gives an estimate of $\pm \varphi \mod M \in [0, M)$ with $\varphi$ (cf. \cref{eq:phi_j}) given by
\begin{equation}
\label{eq:phiarccoslambda}
    \varphi = \frac{M}{2 \pi} \arccos \left(\frac{\lambda}{s \lVert H \rVert_{\mathrm{max}}} \right).
\end{equation}
Accordingly, there exists a $\tilde{\varphi} \in \{0, 1, \dots, 2^{m-1}-1 \}$ such that\footnote{The first bit in the phase register just represents the sign of $\pm \varphi \mod M $, so $\varphi$ is represented with $m-1$ bits.}
\begin{equation}
    |\varphi - \tilde{\varphi} | \le \frac{1}{2},
\end{equation}
to which we can associate an estimated eigenvalue $\tilde{\lambda}$ by inverting \cref{eq:phiarccoslambda}.
The phase stored in the ancilla register corresponds to the estimated $\tilde{\lambda}$ as follows
\begin{equation}
	\tilde{\varphi}
	=
	\frac{M}{2 \pi}\arccos \left( \frac{\tilde{\lambda}}{s \lVert H \rVert_{\mathrm{max}}} \right).
	\label{eq:phase_ev} 
\end{equation}
Thus, we get
\begin{equation}
	\frac{M}{2\pi} \abs{
		\arccos \left( \frac{\lambda}{s \lVert H \rVert_{\mathrm{max}}} \right)
		- \arccos \left( \frac{\tilde{\lambda}}{s \lVert H \rVert_{\mathrm{max}}} \right)
	} \le \frac{1}{2}.
	\label{eq:lim_arccos}
\end{equation}
The mean value theorem for the arc-cosine function states that for $x > \tilde{x}$, $\exists x_0 \in (\tilde{x}, x)$ such that
\begin{equation}
    \frac{|
        \arccos x
        -\arccos\tilde{x}|
    }{
        |x
        -\tilde{x}|
    }
    =
    \frac{1}{\sqrt{1-\left(x_0\right)^2}}.
    \label{eq:mean_value_theorem}
\end{equation}
Applying the mean value theorem to our case we get that $\exists \lambda_0 \in (\tilde{\lambda}, \lambda)$ if $\lambda > \tilde{\lambda}$ or $\exists \lambda_0 \in (\lambda, \tilde{\lambda})$ if $\lambda < \tilde{\lambda}$, such that
\begin{multline}
\label{eq:meanval_arccos}
    s \lVert H \rVert_{\mathrm{max}} \frac{
    \left | \arccos \left( \frac{\lambda}{s \lVert H \rVert_{\mathrm{max}}} \right)
		- \arccos \left( \frac{\tilde{\lambda}}{s \lVert H \rVert_{\mathrm{max}}} \right)\right |}{|\lambda - \tilde{\lambda}|} \\
    = \frac{1}{\sqrt{1 - \left(\frac{\lambda_0}{s \lVert H \rVert_{\mathrm{max}}} \right)^{2}}} \ge 1,
\end{multline}
where the last inequality holds since $0 \le \lambda_0 \le s \lVert H \rVert_{\mathrm{max}} $. Combining \cref{eq:lim_arccos} and \cref{eq:meanval_arccos}, we get
\begin{equation}
	| \lambda - \tilde{\lambda} | \le \frac{\pi s \lVert H \rVert_{\mathrm{max}}}{M},
\end{equation}
which tells us that the error of the estimated eigenvalue is below $\varepsilon$ if the number of qubits in the phase register satisfies
\begin{equation}
	m \ge m_\text{min}^{(1)}
	=
	\left\lceil
		\log_2{\left(
			\frac{\pi s\norm{H}_\text{max}}{\varepsilon}	
		\right)}
	\right\rceil.
 \label{eq:m_min_eval}
\end{equation}

    \subsubsection{Weight approximation}
    \label{subapp:weight_aprx}

We analyze the error on the estimated weights $W_{uj}^2$ of the local response function defined in \cref{eq:weight_apprx} due to the finite size of the phase register. In our analysis we assume that our estimated peak $\tilde{\varphi}_j$ satisfies $\lceil \tilde{\varphi}_j \rceil = \lceil \varphi_j \rceil$ with $\varphi_j$ associated with the true eigenvalue $\lambda_j$ as in \cref{eq:phi_j}.\footnote{An analogous derivation is possible if we assume $\lfloor \tilde{\varphi}_j \rfloor = \lfloor \varphi_j \rfloor$}
Accordingly, our estimator for the weight $W_{uj}^2$ is
\begin{equation}
\label{eq:weightestimate}
    \tilde{W}_{uj}^2
    =
    2\sum_{q=-Q}^{Q-1}
        P(\lceil \varphi_j\rceil+q),
\end{equation}
with the range $Q$ to be determined.

The parameter $Q$ should be taken large enough to guarantee
\begin{equation}
    \abs{
        \tilde{W}_{uj}^2
        -W_{uj}^2
    }
    \leq
    \delta.
\end{equation}
An increase in $Q$ prevents two things.
First, it limits the probability we miss by adding only over the $2Q$ probabilities for values closest to the center of a peak.
This is connected to the lower bound of $\tilde{W}_{uj}^2-W_{uj}^2$.
The second is that we need to increase the distance between two peaks to fit $2Q$ between them which prevents probability ``leakage" between them.
This is connected to the upper bound of $\tilde{W}_{uj}^2-W_{uj}^2$.

For the lower limit we need to study the probability to measure $x$ in the phase register defined in \cref{eq:prob_local} and focus only on one contribution
\begin{equation}
    P(x)
    \geq
    \frac{W_{uj}^2}{2} 
        \abs{a_{x+}^{(j)}}^2.
    \label{eq:one_term_prob}
\end{equation}
We will now further develop $|a_{x+}^{(j)}|^2$ by using its definition \cref{eq:qpe_amp} and the geometric series
\begin{equation}
    \sum_{z=0}^{M-1} x^z 
    =
    \frac{1-x^M}{1-x}.
\end{equation}
Thus, we have
\begin{multline}
    \abs{a_{x+}^{(j)}}^2
    =
    \frac{1}{M^2}
    \abs{
        \frac{
            1-e^{2\pi i (\varphi_j-x)}
        }{
            1-e^{2\pi i (\varphi_j-x)/M}
        }
    }^2
    \\
    =
    \frac{1}{M^2}
    \left[
        \frac{
            \sin(\pi(\varphi_j-x))
        }{
            \sin(\pi(\varphi_j-x)/M)
        }
    \right]^2 \ge \left[
        \frac{
            \sin(\pi(\varphi_j-x))
        }{
            \pi(\varphi_j-x)
        }
    \right]^2,
    \label{eq:a_sin}
\end{multline}
where we used $\abs{\sin x}\leq \abs{x}$. Using \cref{eq:weightestimate}, \cref{eq:one_term_prob} and \cref{eq:a_sin}, we get
\begin{equation}
   \tilde{W}_{uj}^2
    \geq
    W_{uj}^2
    \sum_{q=-Q}^{Q-1}
        \left[
        \frac{
            \sin(\pi(\varphi_j-\lceil \varphi_j\rceil-q))
        }{
            \pi(\varphi_j-\lceil \varphi_j\rceil-q)
        }
    \right]^2,   
\end{equation}
which is minimal for $\lceil \varphi_j\rceil-\varphi_j=1/2$. Hence,
\begin{multline}
    \tilde{W}_{uj}^2
    \geq
    W_{uj}^2
    \sum_{q=-Q}^{Q-1}
        \left[
        \frac{
            1
        }{
            \pi(1/2+q)
        }
    \right]^2 \\ 
    = 2 W_{uj}^2 \sum_{q=1}^Q 
    \left[
        \frac{
            1
        }{
            \pi(q-1/2)
        }
    \right]^2.
    \label{eq:weight_apprx_2}
\end{multline}
Now, we introduce the trigamma function 
\begin{equation}
    \Psi^{(1)}(z) = \sum_{q=0}^{+ \infty} \frac{1}{(z +q)^2},
\end{equation}
which for half integer values takes the form
\begin{equation}
    \Psi^{(1)}\left(Q+\frac{1}{2} \right)
    =
    \frac{\pi^2}{2}
    -
    \sum_{q=1}^{Q}
        \frac{1}{\left(
            q
            -1/2
        \right)^2},
        \label{eq:trigamma}
\end{equation}
with $Q \in \mathbb{N}$. $\Psi^{(1)}(z)$ satisfies
\begin{equation}
    \Psi^{(1)}\left(
        z
    \right)
    \leq
    \frac{z+1}{z^2}
    \leq
    \frac{2}{z},
    \qquad
    \forall z>0.
    \label{eq:lim_trigamma}
\end{equation}
More details about the trigamma function and a derivation of \cref{eq:lim_trigamma} can be found in~\cite{qi2010}.
Combining \cref{eq:weight_apprx_2,eq:trigamma,eq:lim_trigamma} leads to
\begin{multline}
    \tilde{W}_{uj}^2
    \geq
    W_{uj}^2\left(
        1
        - \frac{2}{\pi^2} \psi^{(1)}\left(Q+\frac{1}{2}\right)
    \right)
    \\
    \geq
    W_{uj}^2\left(
        1
        - \frac{4}{\pi^2\left(Q+\frac{1}{2}\right)} 
    \right).
\end{multline}
Since, $W_{uj}^2 \le 1$, this tells us that our estimation $\tilde{W}_{uj}^2$ is at most $\delta$ less than $W_{uj}^2$ if we choose a $Q$ such that
\begin{equation}
    \frac{
        4 
    }{
        \pi^2\left(
            Q
            +1/2
         \right)
    } 
    \leq
    \delta,
    \label{eq:lower_lim_delta}
\end{equation}
or
\begin{equation}
    Q \ge Q_\mathrm{min}^{(1)}
    =\left\lceil
        \frac{4}{\pi^2\delta}
        -\frac{1}{2}
    \right\rceil.
    \label{eq:Qmin1}
\end{equation}

For the upper limit, let us consider 
\begin{multline}
    \tilde{W}_{uj}^2 -W_{uj}^2
    \\
   =
    \sum_{i=0}^{N-1} W_{u i}^2
    \sum_{q=-Q}^{Q-1}
        \left(
              \abs{a_{\lceil \varphi_j\rceil+q,+}^{(i)}}^2
        	   +\abs{a_{\lceil \varphi_j\rceil+q,-}^{(i)}}^2
        \right)
    -W_{uj}^2.
\end{multline}
Since $\sum_{q=-Q}^{Q-1} \abs{a_{\lceil \varphi_j \rceil + q, +}^{(j)}}^2 \leq 1$, we obtain
\begin{multline}
    \tilde{W}_{uj}^2 -W_{uj}^2
    \\
    \leq
    \sum_{i\neq j} W_{u i}^2
    \sum_{q=-Q}^{Q-1}
        \left(
            \abs{a_{\lceil \varphi_j\rceil+q,+}^{(i)}}^2
            +\abs{a_{\lceil \varphi_j\rceil+q,-}^{(i)}}^2
        \right)
    \\
    + W_{uj}^2
    \sum_{q=-Q}^{Q-1}
        \abs{a_{\lceil \varphi_j\rceil+q,-}^{(j)}}^2.
\end{multline}
Let us consider the coefficients $\abs{a_{\lceil \varphi_j \rceil +q, +}^{(i)}}^2$.
We can rewrite them as
\begin{equation}
    \abs{a_{\lceil \varphi_j \rceil +q, +}^{(i)}}^2 = \frac{1}{M^2}
    \left[
        \frac{
            \sin(\pi(\gamma_i +\lceil \varphi_i \rceil -\lceil \varphi_j \rceil-q))
        }{
            \sin(\pi(\gamma_i +\lceil \varphi_i \rceil -\lceil \varphi_j \rceil-q)/M)
        }
    \right]^2,
\end{equation}
with $\gamma_i = \varphi_i - \lceil \varphi_i \rceil \in (-1, 0]$. We assume that $|\lceil \varphi_i \rceil - \lceil \varphi_j \rceil > 2 Q$ $\forall i \neq j$, i.e., that the peaks are separated by at lest $2 Q$. Using the fact that $\abs{a_{\lceil \varphi_j \rceil +q, +}^{(i)}}^2$ decreases by increasing $|\lceil \varphi_i \rceil -\lceil \varphi_j \rceil-q|$ for integer values for any fixed $\gamma_i \in (-1, 0]$, we get
\begin{multline}
    \abs{a_{\lceil \varphi_j \rceil +q, +}^{(i)}}^2 \le \frac{1}{M^2} \left[\frac{\sin(\pi(\gamma_i + Q+1))}{\sin(\pi(\gamma_i + Q+1)/M)}\right]^2
    \\
    =
    |a_{\gamma_i + Q+1}|^2.
    \label{eq:lim_a_gamma}
\end{multline}
The function $|a_{\gamma + Q + 1}|^2$ has exactly one local maximum for $\gamma \in (-1, 0]$, which coincides with the global maximum in the interval. We simply denote by $\bar{\gamma} \in (-1, 0]$ the point where $|a_{\gamma + Q + 1}|^2$ is maximal. The same analysis can be repeated for $\abs{a_{\lceil \varphi_j \rceil +q, -}^{(i)}}^2$. Thus, we get
\begin{multline}
    \tilde{W}_{uj}^2 -W_{uj}^2
    \leq
    \sum_{i\neq j} 
        2W_{ui}^2
        \sum_{q=-Q}^{Q-1}
            \abs{
                a_{\bar{\gamma} + Q + 1}
            }^2 \\
    +W_{uj}^2
    \sum_{q=-Q}^{Q-1}
        \abs{a_{\lceil \varphi_j\rceil+q,-}^{(j)}}^2.
        \label{eq:weight_error_2}
\end{multline}
We will further assume that $Q<\abs{\pm \varphi_j}<M/2-Q$, which guarantees that the two peaks $\varphi_j$ and $-\varphi_j$ are also $2 Q$ away from each other and \cref{eq:lim_a_gamma} is satisfied for $i=j$. 
The previous condition fails only for $\lambda_j \approx s\norm{H}$, where $\pm \varphi_j\approx 0$. This can be prevented by further reducing $H$ with the renormalization factor $\cos(2\pi QM^{-1})$.

Now, we replace the amplitude in the last term of \cref{eq:weight_error_2} as well
and have
\begin{equation}
    \tilde{W}_{uj}^2 -W_{uj}^2
    \leq
    \left(
        \sum_{i\neq j} 
            2W_{ui}^2
        +W_{uj}^2
    \right)
    \sum_{q=-Q}^{Q-1}
        \abs{
            a_{\bar{\gamma} + Q + 1}
        }^2.
\end{equation}
At last, we replace the amplitudes using \cref{eq:a_sin} and use the normalization of the eigenvalues $\sum_i W_{ui}^2=1$ to find
\begin{equation}
    \tilde{W}_{uj}^2 -W_{uj}^2
    \leq
    \frac{4Q}{M^2}\abs{
        \frac{
            \sin(\pi(\bar{\gamma} + Q + 1))
        }{
            \sin(\pi (\bar{\gamma} + Q + 1)/M)
        }
    }^2.
\end{equation}
Now we use $\abs{\sin x }\leq 1$ and $\abs{\sin x}\geq \abs{2x/\pi},\forall \abs{x}\leq \pi/2$ and obtain that our estimation is less than $\delta$ larger than the exact weight $W_{uj}^2$ if
\begin{equation}
    \tilde{W}_{uj}^2 -W_{uj}^2
    \leq
    \frac{Q}{(\bar{\gamma}+ Q + 1)^2}
    \leq
    \frac{1}{Q}
    \leq
    \delta,
    \label{eq:greater_lim_delta}
\end{equation}
if $2Q\leq M-1$.
This yields another lower limit for $Q$, which is
\begin{equation}
    Q^{(2)}_\mathrm{min}
    =
    \left\lceil
        \frac{1}{\delta}
    \right\rceil.
    \label{eq:Qmin2}
\end{equation}
A comparison between \cref{eq:Qmin1,eq:Qmin2} shows that the latter is larger and therefore sufficient to fulfill both conditions. Thus, we set
\begin{equation}
    Q 
    \ge 
    \left \lceil \frac{1}{\delta} \right \rceil.
\end{equation}

The previous analysis depends on the fact that we have a sufficient number of qubits $m$ in the phase register, so that we can distinguish between two neighboring phases. Mathematically, this translates into the condition
\begin{equation}
    \min_i \abs{\varphi_i-\varphi_j}
    \geq
    \min_i \frac{
        M \abs{\lambda_i-\lambda_j} 
    }{
        2\pi s\norm{H}_\text{max}
    }
    \geq 2Q,
\end{equation}
where we used the results obtained in \cref{subapp:err_phase}.
We can guarantee this if we take the number of qubits in the phase register $m$ such that
\begin{equation}
    m \ge m_\text{min}^{(2)}
    =
    \left\lceil
        \log_2\frac{4\pi s\norm{H}_\text{max}}{\delta\Delta_\lambda^{(u)}}
    \right\rceil,
    \label{eq:m_min_weight}
\end{equation}
with $\Delta_{\lambda}^{(u)}$ is defined in \cref{eq:deltalambdau}.
Combining the results of \cref{eq:m_min_eval,eq:m_min_weight}, we obtain 
\begin{multline}
\label{eq:mminmax}
    m \ge \mathrm{max} \left(m_{\mathrm{min}}^{(1)}, m_{\mathrm{min}}^{(2)}  \right)\\
    = \left(\left\lceil
        \log_2\frac{\pi s\norm{H}_\text{max}}{\varepsilon}
    \right\rceil, \left\lceil
        \log_2\frac{4 \pi s\norm{H}_\text{max}}{\delta\Delta_\lambda^{(u)}}
    \right\rceil \right),
\end{multline}
which justifies \cref{eq:ntotscaling} in the main text. 

\subsubsection{Oracle queries for a single run of QPE}
\label{subapp:oraclequeries}
The number of qubits $m$ in the phase register determines the number of oracles queries in the algorithm in \cref{fig:full_phase_estimation}. As we can see from \cref{fig:circ_cqw}, the implementation of a controlled $V$ operator requires to use once $U_T$ and once $U_T^{\dagger}$. The circuit for $U_T$ is shown in \cref{fig:state_trafo} and uses once the position oracle $O_P$, once the angle oracle $O_{\vartheta}$ and once its inverse $O_{\vartheta}^{\dagger}$. Thus, for each $U_T$ we have $3$ oracle calls and a controlled $V$ requires $6$ oracle calls. In order, to perform a single run of QPE with $m$ qubit on the phase register we would need 
\begin{equation}
    \sum_{k=0}^{m-1} 2^k = 2^m -1,
\end{equation}
controlled $V$ operations, which yields the number of oracle queries
\begin{equation}
    N_{\mathrm{queries}} = 6 (2^m -1).
\end{equation}
Using the result of \cref{eq:mminmax} yields the scaling result in \cref{eq:queryscaling} in the main text. 

\subsection{Sample size for given tolerances}
\label{subapp:sample_size}
We now turn our attention to the scaling of the number of samples $N_\mathrm{S}$. Our aim here is to perform a simplified analysis that should capture the asymptotic behavior of the scaling, while the details will eventually depend on the procedure used to infer eigenvalues and weights from measured data in the phase register (see Ref.~\cite{obrien_2019} for instance). In particular, we work with the assumption that the number of qubits $m$ in the phase register is taken large enough to resolve all the peaks. From \cref{eq:m_min_eval} this means 
\begin{equation}
    m \gg \left \lceil \log_2 \frac{\pi s \lVert H \rVert_{\mathrm{max}}}{\Delta_{\lambda}} \right \rceil.
\end{equation}
If this holds we can unambiguously identify the eigenvalue peaks and neglect possible errors in the assignment of the samples to the eigenvalues. In this regime, QPE de facto behaves as a measurement in the eigenbasis of $H$, where each eigenvalue $\lambda_j$ appears with probability $p(j) = W_{uj}^2$ \footnote{In our version of QPE we have two peaks associated with each eigenvalue, but we can simply sum the probability of each peak to get $W_{uj}^2$}. By performing QPE many times, we obtain samples from which we can reconstruct the probability distribution $W_{uj}^2$.

Let $\Lambda_u$ be the set of eigenvalues $\lambda_j$ that have ``support" on the oscillator $u$, i.e., 
\begin{equation}
    \Lambda_u = \{ \lambda_j : \, W_{uj}^2 > 0 \},
\end{equation}
and $N_u = |\Lambda_u |$ the number of elements of $\Lambda_u$ ($N_u \le N$).
$W_{uj}^2$ is a probability distribution over the finite alphabet $\Lambda_u$. 

Let $p(\lambda)$ be a generic probability distribution over the finite alphabet $\Lambda_u$. With each $\lambda_j \in \Lambda_u$ we can associate a Bernoulli random variable $X_{\lambda_j}(\lambda)$ defined $\forall \lambda \in \Lambda_u$ as
\begin{equation}
    X_{\lambda_j}(\lambda) = \begin{cases}
        1, \quad \lambda=\lambda_j, \\
        0, \quad \lambda \neq \lambda_j.
    \end{cases}
\end{equation}
Notice that $\mathrm{E}[X_{\lambda_j}]=p(\lambda_j)$. By drawing $N_\mathrm{S}$ samples from $p(\lambda)$, we get samples $x_{\lambda_j}^{(i)}$, $i\in\{1, \dots, N_\mathrm{S}\}$ for $X_{\lambda_j}$. By taking the arithmetic average of these samples we can construct the empirical estimator for $p(\lambda_j)$:
\begin{equation}
    \tilde{p}(\lambda_j) = \frac{1}{N_\mathrm{S}} \sum_{i=1}^{N_\mathrm{S}} x_{\lambda_j}^{(i)}.
\end{equation}
Our aim is to understand how many samples $N_\mathrm{S}$ we have to draw from $p(\lambda)$ to have, with probability $1-\zeta$, that each $\tilde{p}(\lambda_j)$, $\lambda_j \in \Lambda_u$ estimated $p(\lambda_j)$ with accuracy $\delta$. Mathematically, this means
\begin{equation}
\label{eq:probwujdelta}
    \mathrm{Prob} \left( \bigcup_{\lambda_j \in \Lambda_u} \{
        |p(\lambda_j) - \tilde{p}(\lambda_j) | 
        \ge 
        \delta 
    \}\right) 
    \le 
    \zeta.     
\end{equation}
Thus, our aim is to obtain bounds for the left-hand side of \cref{eq:probwujdelta}. First, using the union bound we get
\begin{multline}
\label{eq:ljunion}
    \mathrm{Prob} \left( \bigcup_{\lambda_j \in \Lambda_u} \{
        |p(\lambda_j) - \tilde{p}(\lambda_j) | 
        \ge 
        \delta
    \}
    \right) \\
    \le 
    \sum_{\lambda_j \in \Lambda_u} \mathrm{Prob}\left(
        |p(\lambda_j) - \tilde{p}(\lambda_j)| 
        \ge 
        \delta 
    \right).
\end{multline}
Now, an application of Hoeffding's inequality \cite{hoeffding1963, DemboZeitouni} to each random variable $X_{\lambda_j}$ yields
\begin{multline}
\label{eq:hoeffxj}
\mathrm{Prob}\left(|p(\lambda_j) - \tilde{p}(\lambda_j)| \ge \delta \right) \\
= \mathrm{Prob}\left(\left \lvert \mathrm{E}[X_{\lambda_j}] - \frac{1}{N_\mathrm{S}} \sum_{i=1}^{N_\mathrm{S}} x_{\lambda_j}^{(i)} \right \rvert \ge \delta \right) \le 2 e^{-2 N_\mathrm{S} \delta^2}.
\end{multline}
Plugging \cref{eq:hoeffxj} into \cref{eq:ljunion} and enforcing \cref{eq:probwujdelta} we get
\begin{multline}
    \mathrm{Prob} \left( \bigcup_{\lambda_j \in \Lambda_u}\{ |p(\lambda_j) - \tilde{p}(\lambda_j) | \ge \delta\} \right) \\
    \le 2 N_u e^{-2 N_\mathrm{S} \delta^2} \le \zeta,
\end{multline}
which holds if
\begin{equation}
\label{eq:nsapp}
    N_\mathrm{S} \ge \frac{1}{2 \delta^2} \ln\left(\frac{2 N_u}{\zeta} \right).
\end{equation}
\cref{eq:nsapp} justifies \cref{eq:samplescaling} in the main text. 


\section{Modification for nonlocal response}
\label{app:nonlocal_resp}
In this appendix, we describe how to modify our algorithm to compute nonlocal entries $G_{uv}(\lap)$ of the response function whose analytical form is given in \cref{eq:guv}.
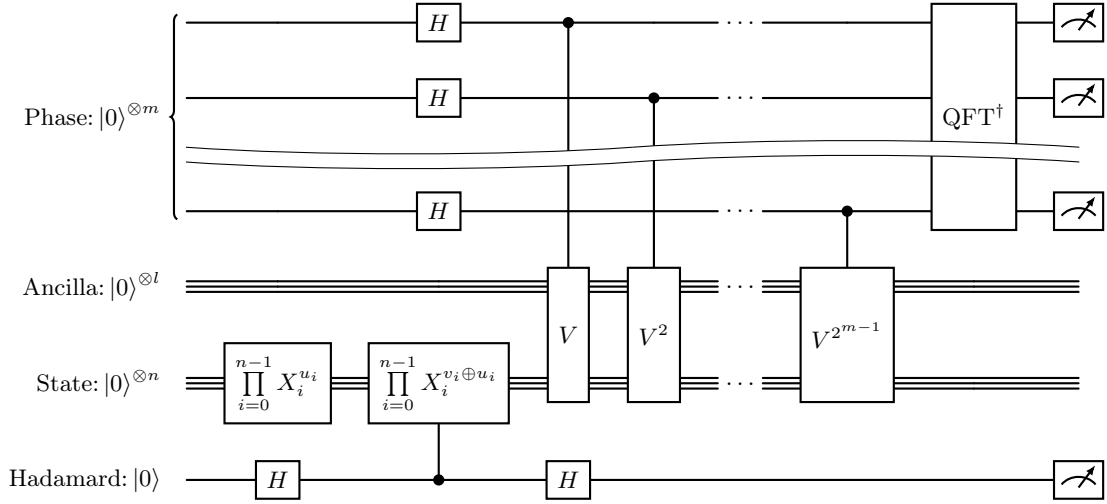
\begin{figure*}
	\centering
	\begin{adjustbox}{max width=0.99\linewidth}
		\begin{quantikz}[wire types= {q,q,q,q,b,b},classical gap=0.07cm]
			\lstick[wires=4, label style={anchor= east, xshift=-0cm}]{$\text{Phase:}\ket{0}^{\otimes m}$} &&
            \gate{H}&   
            \ctrl{5}&& \ \ldots\ && 
            \gate[4,disable auto height]{\text{QFT}^\dagger}
            &\meter{} 
			\\
			&& \gate{H}&& \ctrl{4}& \ \ldots\ &&&\meter{}
			\\
			&\wave&&&&&&&&&&&
			\\
			&& \gate{H}&&&\ \ldots\ & \ctrl{2}&&\meter{}
			\\
			\lstick[label style={anchor= east, xshift=-0.2cm}]{$\text{Ancilla:}\ket{0}^{\otimes l}$}
            & &
            & \gate[2, disable auto height]{V}&\gate[2, disable auto height]{V^2}&\ \ldots\ & \gate[2,disable auto height]{V^{2^{m-1}}}&&
			\\
			\lstick[label style={anchor= east, xshift = -0.2cm }]{$\text{State:}\ket{0}^{\otimes n}$}& 
            \gate{\prod\limits_{i=0}^{n-1}X_i^{u_i}} & 
            \gate{\prod\limits_{i=0}^{n-1}X_i^{v_i\oplus u_i}}&&& \ \ldots\ &&&
            \\
            \lstick[label style={anchor= east, xshift=-0.2cm}]{$\text{Hadamard:}\ket{0}$} &\gate{H}&\ctrl{-1}&\gate{H}&&&&&\meter{}
		\end{quantikz}
	\end{adjustbox}
	\caption{
        Modified circuit for the computation of the nonlocal response function. 
        One can see two additions compared to \cref{fig:full_phase_estimation}.
        The first is an additional qubit and the second is a modified Hadamard test which we use to prepare the state $\ket{u}$ next to $\ket{v}$.
    }
	\label{fig:nonlocal_phase_estimation}
\end{figure*}
For the local contributions the preparation of the computational basis state $\ket{u}$ is crucial to extract the weights $W_{uj}^2$.
The disadvantage of this method is that we can only extract the absolute square of those weights, which does not contain the relative sign and makes this method not suited to evaluate the terms $W_{uj}W_{vj}$ that appear in \cref{eq:guv}.

A way around this is to add another qubit and still prepare $\ket{u}$ in the state register as shown in \cref{fig:nonlocal_phase_estimation}.
We will now use the modified Hadamard test to ``add" $\ket{v}$ to this state.
At first we apply a Hadamard gate to the additional ancilla qubit starting in $\ket{0}$ (bottom qubit in \cref{fig:nonlocal_phase_estimation}). 
Thus, we have
\begin{equation}
   \xrightarrow{H} \frac{1}{\sqrt{2}}(\ket{0}+\ket{1})\ket{u}.
\end{equation}
We now apply a series of CNOT gates with the ancilla qubit as control and the qubits in the $n$-qubit register as target. 
The CNOT will be only applied to target qubit $i$ for which $v_i \oplus u_i =1$. 
This brings the state $\ket{u}$ to $\ket{v}$ when the control qubit is in $\ket{1}$. Accordingly, after this series of CNOTs we get
\begin{equation}
    \xrightarrow{\mathrm{CNOTs}} \frac{1}{\sqrt{2}} (
        \ket{0}\ket{u}
        +\ket{1}\ket{v}
    ).
\end{equation}
At last we apply another Hadamard gate to the ancilla qubit and obtain 
\begin{equation}
   \xrightarrow{H} \frac{1}{2}\left(
        \ket{0}\left(\ket{u}+\ket{v}\right)
        +\ket{1}\left(\ket{u}-\ket{v}\right)
    \right).
\end{equation}
Measuring the additional qubit results in either the sum or difference of those two states
\begin{equation}
    \frac{1}{\sqrt{2}}\left(
        \ket{u}\pm\ket{v}
    \right)
    =
    \frac{1}{\sqrt{2}}\sum_{j=1}^N \left(
        W_{uj}
        \pm W_{vj}
    \right) \ket{\lambda_j}.
\end{equation}
If we proceed now as before with the local version of the algorithm, depending on the ancilla being in $0$ or $1$, we would extract the square of those amplitudes
\begin{equation}
    \frac{1}{2}(W_{uj}\pm W_{vj})^2
    =
    \frac{1}{2}W_{uj}^2
    +\frac{1}{2}W_{vj}^2
    \pm W_{uj}W_{vj},
\end{equation}
where we used that $W_{uj}$ and $W_{vj}$ are real. Let $P(0, x)$, $P(1, x)$ the probability of measuring the bitstring $x$ in the phase register given that the ancilla is measured in $0$ and $1$, respectively. 
By adapting \cref{eq:weight_apprx} with the new amplitudes we find
\begin{subequations}
\begin{equation}
    \frac{1}{2}(W_{vj} + W_{uj})^2
    \approx
    2\sum_{q=-Q}^{Q-1} P(0,\lceil \varphi_j'\rceil+q),
\end{equation}
\begin{equation}
    \frac{1}{2}(W_{vj} - W_{uj})^2
    \approx
    2\sum_{q=-Q}^{Q-1} P(1,\lceil \varphi_j'\rceil+q),
\end{equation}
\end{subequations}
or, equivalently,
\begin{equation}
    W_{vj}W_{uj}
    \approx
    \sum_{q=-Q}^{Q-1} \left(
        P(0,\lceil \varphi_j'\rceil+q)
        - P(1,\lceil \varphi_j'\rceil+q)
    \right).
\end{equation}

\section{Details of the glued-trees problem}
\label{app:gluedtree}
In this appendix, we provide the necessary details of the glued-trees problem. We consider a random glued-tree graph $\mathcal{G}=(\mathcal{V}, \mathcal{E})$ shown in \cref{fig:randomglued} and described in \cref{sec:glued_tree}. We work directly with the quantum formalism so that, with each vertex $u \in \mathcal{V}$, we associate an element in the computational basis $\ket{u}$. 

First, we note that, by definition, the action of an adjacency matrix $A$ on a vertex $\ket{u}$ reads
\begin{equation}
    A \ket{u}= \sum_{v \in N_\mathcal{G}[u] \setminus \{u\}} \ket{v}.
\end{equation}
As we see from \cref{fig:randomglued} the vertices in the glued-trees graph can be organized in $2 n_c$ columns. 
Let us denote by $\mathrm{col}(j)$ the set of vertices in the $j$-th columns starting from the left of the random glued tree graph with $j \in\{ 1, \dots , 2 n_c\} $. We define the ``column" states

\begin{subequations}
\begin{equation}
\ket{\mathrm{col}(j)} 
= 
\frac{1}{\sqrt{2^{j - 1}}} \sum_{v \in \mathrm{col}(j)} \ket{v}, \quad j\in\{1, \dots , n_c\},
\end{equation}
\begin{equation}
\ket{\mathrm{col}(j)} = \frac{1}{\sqrt{2^{2 n_c -j}}} \sum_{v \in \mathrm{col}(j)} \ket{v}, \quad j\in\{n_c + 1, \dots , 2 n_c\}.
\end{equation}
\end{subequations}
Notice that ENTRANCE and EXIT vertices correspond to $\ket{\mathrm{col}(1)}$ and $\ket{\mathrm{col}(2 n_c)}$, respectively. 
We denote by $\mathcal{H}_{\mathrm{col}}$ the column subspace:
\begin{equation}
    \mathcal{H}_{\mathrm{col}} 
    = 
    \mathrm{span}\{ 
        \ket{\mathrm{col}(j) }: j\in\{1, \dots, 2 n_c\}
    \}.
\end{equation}
The projector onto $\mathcal{H}_{\mathrm{col}}$ is
\begin{equation}
\Pi_{\mathrm{col}} = \sum_{j=1}^{2 n_c} \ketbra{\mathrm{col}(j)}{\mathrm{col}(j)}. 
\end{equation}
The crucial fact is that if the system starts in the column subspace, an application of the adjacency matrix $A$ will give a state in the column subspace. 
In fact, the action of the adjacency matrix $A$ of the random glued-trees graph on the column states reads
\begin{equation}
A \ket{\mathrm{col}(j)} = \sqrt{2} \ket{\mathrm{col}(j+1)} + \sqrt{2} \ket{\mathrm{col}(j-1)},
\end{equation}
for $j \in\{2, \dots, n_c -1\}\cup\{ n_c +1, \dots 2 n_c -1\}$, and
\begin{subequations}
\begin{equation}
A \ket{\mathrm{col}(1)} = \sqrt{2} \ket{\mathrm{col}(2)}, 
\end{equation}
\begin{equation}
A \ket{\mathrm{col}(2 n_c)} = \sqrt{2} \ket{\mathrm{col}(2 n_c -1)}, 
\end{equation}
\begin{equation}
A \ket{\mathrm{col}(n_c)} = \sqrt{2} \ket{\mathrm{col}(n_c -1)} + 2 \ket{\mathrm{col}(n_c +1)}, 
\end{equation}
\begin{equation}
A \ket{\mathrm{col}(n_c +1)} = 2 \ket{\mathrm{col}(n_c)} + \sqrt{2} \ket{\mathrm{col}(n_c +2)}.
\end{equation}
\end{subequations}
For notational simplicity we will denote the column states $\ket{\mathrm{col}(j)}$ as $\ket{j}$ from now on:
\begin{equation}
\ket{\mathrm{col}(j)} \mapsto \ket{j}.
\end{equation} 

Since in \cref{prob:gluedtreeprob} the system starts at the ENTRANCE, i.e., in $\ket{\mathrm{ENTRANCE}}=\ket{j=1}$, we can effectively restrict to the column subspace and consider the projected adjacency matrix 
\begin{multline}
\label{eq:api}
\frac{A_{\mathrm{col}}}{\sqrt{2}}
= 
\frac{\Pi_{\mathrm{col}} A \Pi_{\mathrm{col}}}{\sqrt{2}} 
= 
\sum_{j=1}^{n_c -1}(\ketbra{j}{j +1} + \mathrm{h.c.}) \\
 + \sqrt{2}(\ketbra{n_c}{n_c+1}  +  \mathrm{h.c.}) \\
 + 
\sum_{j=n_c+1}^{2 n_c -1}(\ketbra{j}{j +1} + \mathrm{h.c.})  ,
\end{multline}
as the effective Hamiltonian of the system. Eq.~\eqref{eq:api} is the Hamiltonian of a quantum walk on one line with $2 n_c$ sites with a ``defect" in the middle. Note that from a matrix point of view $A_{\mathrm{col}}$ is a tridiagonal symmetric matrix. 

As shown in Ref.~\cite{childs2003},  $A_{\mathrm{col}}/\sqrt{2}$ has eigenvalues of the form
\begin{equation}
    \lambda_z = 2  \cos z,
\end{equation}
where $z=p+i q \in \mathbb{C}$, $p, q \in \mathbb{R}$, satisfies the quantization condition
\begin{equation}
\label{eq:quantcond}
\frac{\sin(z(n_c +1))}{\sin(z n_c)} = \pm \sqrt{2}.
\end{equation}
The corresponding eigenvector is of the form
\begin{multline}
\label{eq:eigansatz}
\ket{\lambda_z} = \frac{1}{\sqrt{\mathcal{N}(z, n_c)}}\biggl(\sum_{j=1}^{n_c} \sin(z j) \ket{j} \\
\pm \sum_{j=n_c +1}^{2 n_c} \sin(z (2 n_c +1 -j)) \ket{j} \biggr),
\end{multline} 
with $\mathcal{N}(z, n_c)$ a normalization factor. 
\begin{figure}
\centering
\includegraphics[height=6cm]{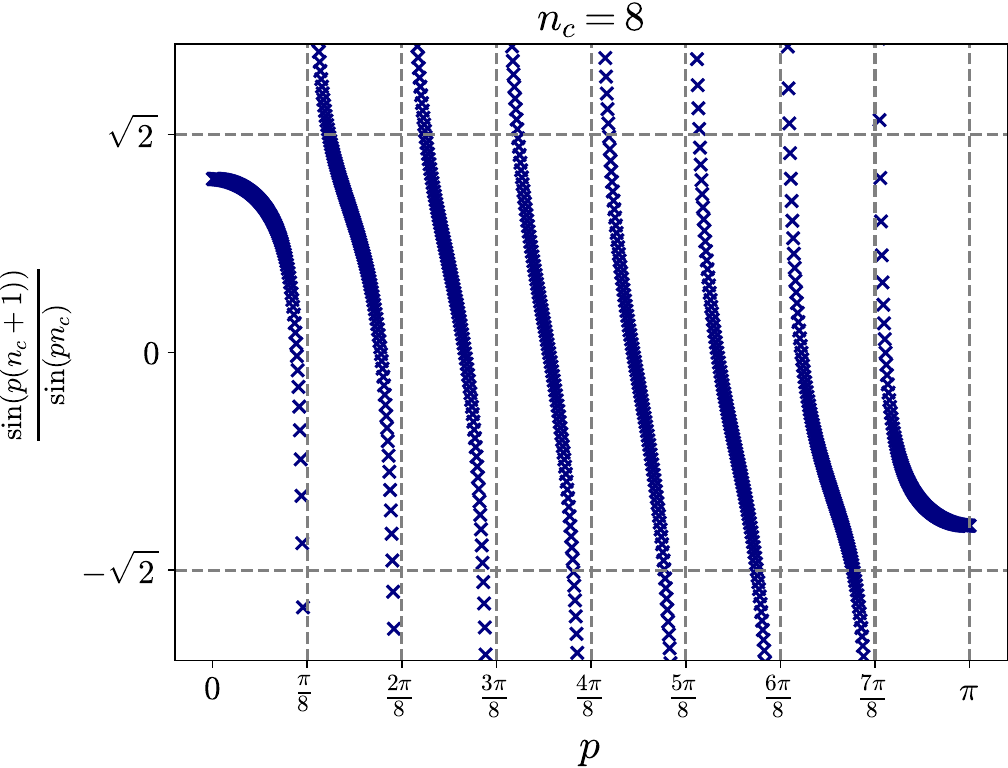}
\caption{Quantization condition \cref{eq:quantcond} for $n_c=8$.}
\label{fig:quantcond}
\end{figure}
We denote by $\mathcal{Z}$ the set of solutions of \cref{eq:quantcond} that are associated with a distinct eigenpair. Thus, $| \mathcal{Z} | =2 n_c$. The quantization condition \cref{eq:quantcond} has exactly $2 n_c -2$ real solutions $z=p$ (see \cref{fig:quantcond} for an example). We denote by $\mathcal{P}$ the set of these solutions. The two remaining eigenpairs can be obtained by setting $z=i q$ and $z= \pi + i q$, but we do not need them to show our result.

In terms of the eigenvectors $\ket{\lambda_z}$, the $\ket{\mathrm{ENTRANCE}}$ state reads
\begin{equation}
    \ket{\mathrm{ENTRANCE}} = \sum_{z \in \mathcal{Z}} \frac{\sin z}{\sqrt{\mathcal{N}(z, n_c)}}  \ket{\lambda_z}.
\end{equation}
We now assume that the number of qubits in the phase register is taken to satisfy \cref{eq:mglued} in the main text. With this assumption, the probability of obtaining the eigenstate $\ket{\lambda_z}$ after applying QPE to $\ket{\mathrm{ENTRANCE}}$, as described in \cref{sec:algo}, and measurement on the ancilla register with post-selection on the $\ket{0}^{\otimes a}$ state, is
\begin{equation}
    \mathrm{Prob}(\lambda_z| \mathrm{ENTRANCE} ) = \frac{|\sin z|^2}{2 \mathcal{N}(z, n_c)}.
\end{equation}
Once we obtained one of the $\ket{\lambda_z}$, the probability of finding the EXIT after measurement in the computational basis is
    \begin{equation}
      \mathrm{Prob}(\mathrm{EXIT}|\lambda_z ) = \frac{|\sin z|^2}{\mathcal{N}(z, n_c)},  
    \end{equation}
Thus, the total probability of finding the EXIT is
\begin{equation}
    \mathrm{Prob}(\mathrm{EXIT}) = \sum_{z \in \mathcal{Z}} \frac{|\sin z|^4}{2 \mathcal{N}(z, n_c)^2}.
\end{equation}

We now want to study analytically the scaling of this probability $\mathrm{Prob}(\mathrm{EXIT})$. First of all, we restrict the sum over the real solutions $z=p$ of \cref{eq:quantcond}. In this case, the normalization factor $\mathcal{N}(p, n_c)$ reads 
\begin{multline}
\label{eq:normfactorep}
    \mathcal{N}(p, n_c) = \sum_{j=1}^{n_c} \sin^2(p j) + \sum_{j=n_c +1}^{2 n_c} \sin^2(p(2 n_c +1 -j)) \\
    = \frac{1}{2}\biggl[1 + 2 n_c - \frac{\sin(p(1 + 2 n_c))}{\sin p} \biggr] < 2 n_c.
\end{multline}
Thus, we get
\begin{equation}
    \mathrm{Prob}(\mathrm{EXIT}) >  \sum_{p \in \mathcal{P}} \frac{\sin^4 p}{2 \mathcal{N}(p, n_c)^2} > \frac{1}{8 n_c^2} \sum_{p \in \mathcal{P}} \sin^4 p.
\end{equation}
As discussed in Ref.~\cite{childs2003} the real solutions concentrate around $\pi \ell/n_c$ with $\ell \in \{1, \dots, n_c -1\}$, with corrections of order $\mathcal{O}(1/n_c^2)$. This is also visible in the example shown in \cref{fig:quantcond} for $n_c=8$. In particular, we see that there are two solutions around $\pi \ell/n_c$. Thus, for large $n_c$, we can write
\begin{multline}
    \sum_{p \in \mathcal{P}} \sin^4 p = 2 \sum_{\ell =1}^{n_c-1} \sin^4 \left(\frac{\pi \ell}{n_c}\right) + \mathcal{O}(1) \\
    = \frac{3 n_c}{4} + \mathcal{O}(1).
\end{multline}
We finally obtain
\begin{equation}
\label{eq:lowerboundpexit}
    \mathrm{Prob}(\mathrm{EXIT}) > \frac{3}{32 n_c} + \mathcal{O}\left(\frac{1}{n_c^2} \right),
\end{equation}
which justifies \cref{eq:gluedscaling} in the main text. 
\end{document}